\documentstyle[preprint]{aastex}

\def  \be	 {\begin{equation}}
\def  \ee	 {\end{equation}}
\def  \beq	 {\begin{eqnarray}}
\def  \eeq	 {\end{eqnarray}}
\def  \simlt	 {\lesssim}
\def  \simgt	 {\gtrsim}
\def  \rlarrows	 {{\lower.6ex\hbox{$\;\buildrel{\mbox{\footnotesize$\longrightarrow$}}
		 \over{\mbox{\footnotesize$\longleftarrow$}}\;$}}}

\def  \define	 {\stackrel{\rm def}{=}}

\begin{document}

\title{\bf Transport phenomena in stochastic magnetic mirrors}

\author{Leonid Malyshkin$^1$ and Russell Kulsrud$^2$}

\affil{Princeton University Observatory, Princeton NJ 08544, USA}
\email{$^1$leonmal@astro.princeton.edu, $^2$rkulsrud@astro.princeton.edu}


\date{\today}


\begin{abstract}
Parallel thermal conduction along stochastic magnetic field lines may be reduced because the heat 
conducting electrons become trapped and detrapped between regions of strong magnetic field 
(magnetic mirrors). The problem reduces to a simple but realistic model for diffusion of mono-energetic 
electrons based on the fact that when there is a reduction of diffusion, it is controlled by a subset of 
the mirrors, the principle mirrors.
The diffusion reduction can be considered as equivalent to an enhancement of the pitch angle 
scattering rate. Therefore, in deriving the collision integral, we modify the pitch angle scattering 
term. We take into account the full perturbed electron-electron collision integral, as well as the 
electron-proton collision term.
Finally, we obtain the four plasma transport coefficients and the effective thermal conductivity.
We express them as reductions from the classical values. We present these reductions as functions of 
the ratio of the magnetic field decorrelation length to the electron mean free path at the thermal 
speed $V_T=\sqrt{2kT/m_e}$. 
We briefly discuss an application of our results to clusters of galaxies.
\end{abstract}



\section{Introduction}\label{INTRODUCTION}

The problem of thermal conduction in a stochastic magnetic field is crucial for our understanding of 
galaxy cluster formation (Suginohara \& Ostriker 1998; Cen \& Ostriker 1999) and for the theory of 
cooling flows (Fabian 1990).
It is also of great interest for the solar physics and for various questions of plasma physics. 
At the same time, the question: ``whether electron thermal conduction is so strongly inhibited by a
stochastic magnetic field in a galaxy cluster, that it can be neglected'', is a very controversial one
(Rosner \& Tucker 1989; Tribble 1989; Tao 1995; Pistinner \& Shaviv 1996; Chandran \& Cowley 1998).
It is currently estimated that if the coefficient of thermal conductivity is less than $1/30$ of the 
Spitzer value, then the time scale of the heat conduction in the cluster is more than the Hubble time 
(Suginohara \& Ostriker 1998). Otherwise, thermal conduction is 
important\label{FOOTNOTE}.~\footnote{This numerical estimate, $1/30$ of the Spitzer value, is based on 
numerical simulations with limited resolution, so it is not the last word on the problem.}

The problem of thermal diffusion of heat conducting electrons in a stochastic magnetic field should 
be divided into two separate parts because there are two separate effects that reduce diffusion 
in the presence of stochastic magnetic field (Pistinner \& Shaviv 1996; Chandran, Cowley, \& Ivanushkina 1999).
The first effect is that the heat conducting electrons have to travel along tangled magnetic 
field lines, and as a result, they have to go larger distances between hot and cold regions of 
space. (In other words, the temperature gradients are weaker along magnetic field lines.) 
The second effect is that electrons, while they are traveling along the field lines, become trapped and 
detrapped between magnetic mirrors (which are regions of strong magnetic field). A trapped electron 
is reflected back and forth between magnetic mirrors until collisions make its pitch angle sufficiently 
small for the electron to escape the magnetic trap. 

In this paper we concentrate on the second effect, and we derive the reduction of the effective electron 
thermal conduction parallel to the magnetic field lines caused by the presence of stochastic magnetic 
mirrors. 

As is well known, a temperature gradient produces electrical current as well as heat flow. Similarly, 
an electric field produces heat flow as well as current. The four transport coefficients describing this
are given in equation~(\ref{J}) and~(\ref{Q}). The transport coefficients were first calculated by
Spitzer \& H$\ddot{\rm a}$rm for an unmagnetized plasma (Spitzer \& H$\ddot{\rm a}$rm 1953; 
Cohen, Spitzer \& Routly 1950). Their coefficients also apply in an uniform magnetic field
for transport parallel to the field. In this paper, we show how the parallel transport coefficients can 
be reduced in the presence of stochastic magnetic mirrors, and we calculate their reduced values
by the same kinetic approach as that of Spitzer \& H$\ddot{\rm a}$rm. The reduction factors are 
presented in Figure~\ref{FIG_TRANSPORT}. The reduced effective thermal conductivity (that resulting
when the electric field is present to cancel the current) is given in Figure~\ref{FIG_CONDUCTION}.
Spatial diffusivity of mono-energetic electrons along the magnetic field lines is also presented 
in Figure~\ref{FIG_D/D_0}.

First, in Section~\ref{ESCAPE}, we solve the kinetic equation to find the escape time $\tau_m$ for 
electrons trapped between two equal magnetic mirrors. We assume, that all electrons have a single 
value of speed, $V$, i.e.~they are mono-energetic. The exact calculations of the escape time are 
given in Appendices~\ref{A_CASE_1} and~\ref{A_CASES_2_3}. In addition, we carry out Monte-Carlo
particle simulations to confirm our results.

Second, in Section~\ref{DIFFUSION}, we apply our results for this escape time to find the reduction of
diffusion of mono-energetic electrons in a system of stochastic mirrors. 
It turns out that in the limit $l_0\gg\lambda$, where $l_0$ is the magnetic field decorrelation length 
and $\lambda$ is the electron mean free path, the parallel diffusivity is unaffected by magnetic mirrors, 
and is given by the standard value $D_0=(1/3)V\lambda$. 
In the opposite limit, $l_0\ll\lambda$, magnetic mirrors do reduce diffusivity. We find that in this 
case there is a subset of the mirrors, the principle mirrors, that inhibits diffusion the most. These 
are mirrors whose separation distances are approximately equal to the electron {\it effective mean free path},
$\lambda_{\rm eff}$, the typical distance that electrons travel in the loss cones before they are scattered 
out of them. In order to estimate the reduction of diffusion in this limit, we need consider only the 
principle mirrors, neglecting all others. Again, we perform the numerical simulations to support these 
theoretical results.

Third, in Section~\ref{EQUATION}, in order to carry out a precise kinetic treatment involving all electrons,
we consider the diffusion reduction to be equivalent to an enhancement 
of the pitch angle scattering rate of electrons. In deriving the collision integral, we, therefore, modify 
the pitch angle scattering term by the inverse of the factor by which the spatial diffusion is reduced.
We take into account the full perturbed electron-electron collision integral, as well as the electron-proton 
collision term. We obtain an integro-differential equation for the perturbed electron distribution function 
in the presence of stochastic magnetic mirrors. If there is no reduction of electron diffusivity, 
our equation reduces to the well known result obtained by Spitzer and H$\ddot{\rm a}$rm 
(Spitzer \& H$\ddot{\rm a}$rm 1953; Cohen, Spitzer \& Routly 1950; Spitzer 1962).

Fourth, in Section~\ref{TRANSPORT}, we solve our equation numerically, separately for the Lorentz gas 
in the presence of magnetic mirrors, neglecting electron-electron collisions (in this case the equation
simplifies greatly), and for the Spitzer gas in the presence of magnetic mirrors. We find the reductions of 
the four plasma transport coefficients and of the effective thermal conductivity as functions of the ratio 
of the magnetic field decorrelation length $l_0$ to the electron mean free path at the thermal speed 
$V_T=\sqrt{2kT/m_e}$ (this mean free path is different for the Lorentz and Spitzer models). 
We find that the major effect of the magnetic mirrors is the reduction of anisotropy of superthermal 
electrons (this anisotropy is driven by a temperature gradient or/and by an electric field). Electrical 
current and heat are mainly transported by these electrons, whose diffusivity is suppressed the most.

Finally, we discuss our results and give the conclusions in Section~\ref{CONCLUSIONS}.


\section{Mono-energetic electrons trapped between two equal magnetic mirrors}\label{ESCAPE}

\begin{figure}[t]
\vspace{6.4cm}
\includegraphics{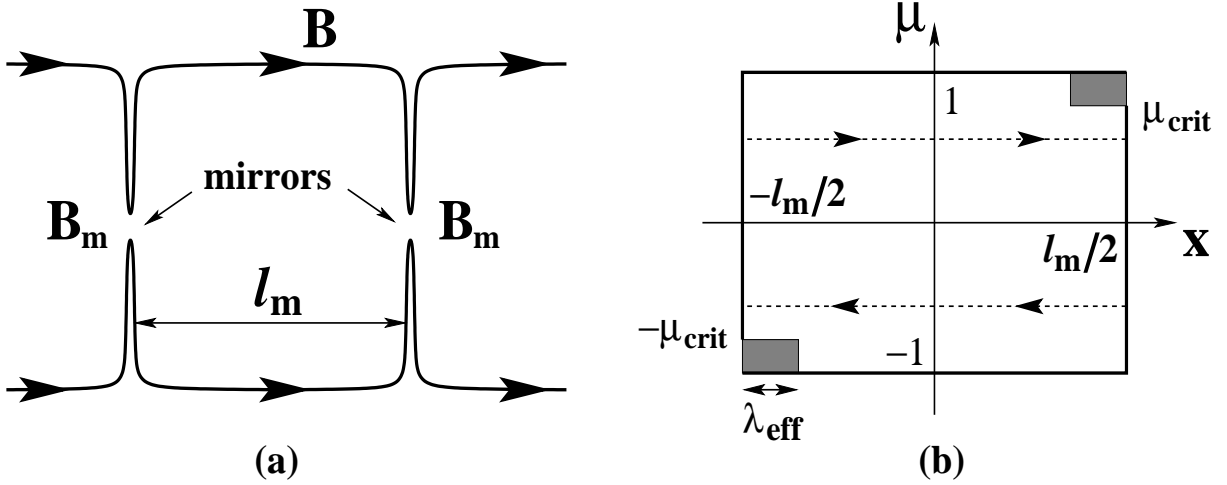}
\caption{(a): A magnetic flux tube with two ``step-function like'' magnetic mirrors. The mirror strengths are 
$m=B_{\rm m}/B$. (b): The phase space box where electrons are trapped in coordinates $x$ and $\mu=\cos\theta$. 
The horizontal dotted lines show a closed trajectory of a trapped electron in the limit $l_m\ll\lambda$. 
The electrons escape the magnetic trap through two escape windows: $x=l_m/2$, $\mu>\mu_{\rm crit}=\sqrt{1-1/m}$ 
and $x=-l_m/2$, $\mu<-\mu_{\rm crit}$. In the limit $l_m\ll \lambda_{\rm eff}$ the electrons freely escape 
to the right or left whenever they reach the two loss cones, $\mu>\mu_{\rm crit}$ and $\mu<-\mu_{\rm crit}$. 
In the opposite limit, $\lambda_{\rm eff}\ll l_m$, the electrons escape when they reach the two shaded regions 
of the phase space.}
\label{FIG_MIRRORS}
\end{figure}

In this section we solve the kinetic equation to find the escape time $\tau_m$ for electrons trapped 
between two equal magnetic mirrors. We assume here and in the next section that all electrons have a 
single value of speed, $V$, which is unchanged by collisions, i.e.~electrons are mono-energetic.
In order to derive an analytical solution, we make several additional simplifying 
assumptions. Let the two magnetic barriers (mirrors) be both equal to $B_{\rm m}$, and we assume the magnetic 
field $B$ is constant between them. We introduce the {\it mirror strength} $m\define B_{\rm m}/B$. The 
separation of the mirrors is $l_m$, and their thicknesses are negligible compared to $l_m$. In other words, 
magnetic mirrors are similar to thin step-functions with heights $B_{\rm m}-B$ and with constant field $B$ 
between them (see Figure~\ref{FIG_MIRRORS}). This is a reasonable assumption, because as we will see in the 
next section, electron diffusion is controlled by strong mirrors with mirror strengths $m\simgt 4$, which 
are separated by distances much larger than the magnetic field decorrelation length (if the spectrum of mirrors 
falls off with their strength significantly faster than $1/m$, the case that we consider in this paper).

Under these assumptions, the kinetic equation for the distribution function $f(t,x,\mu)$ of mono-energetic 
electrons trapped between the two mirrors is (Braginskii 1965)
\beq
\frac{\partial f}{\partial t}+\mu V\frac{\partial f}{\partial x}=
\frac{\nu}{2}\frac{\partial}{\partial\mu}\left[(1-\mu^2)\frac{\partial f}{\partial\mu}\right].
\label{FULL_KINETIC_EQ}
\eeq
Here $x$ is one-dimensional space coordinate along a magnetic flux tube, $t$ is time, $\mu=\cos\theta$ is
the cosine of the electron's pitch angle, and $\nu=V/\lambda$ is the collision frequency [$\lambda$ is
the mean free path, see equations~(\ref{LAMBDA_L_T}) and~(\ref{LAMBDA_S_T})]. The right-hand side of 
equation~(\ref{FULL_KINETIC_EQ}) represents the pitch angle scattering rate, $\nu$, of electrons.
The electrons are trapped in the region of space between the mirrors, $-l_m/2<x<l_m/2$, and they can escape 
through the two windows: $x=l_m/2$, $\mu>\mu_{\rm crit}=\sqrt{1-1/m}$ and $x=-l_m/2$, $\mu<-\mu_{\rm crit}$, 
as shown in Figure~\ref{FIG_MIRRORS}. 
The mirror strength is $m=B_{\rm m}/B$, and it is the measure of the relative heights of the magnetic barriers. 
For simplicity, we assume that the barriers are high, i.e. $m\gg 1$ and $\mu_{\rm crit}\approx 1-1/2m$. 
In this case the electron distribution is in quasi-static equilibrium,
\beq
f(t,x,\mu)=e^{-t/\tau_m}F(x,\mu), \quad\tau_m\gg\nu^{-1},
\label{F_TIME_DEPENDENT}
\eeq
and equation~(\ref{FULL_KINETIC_EQ}) reduces to 
\beq
-\frac{F}{\tau_m}+\mu V\frac{\partial F}{\partial x}=
\frac{\nu}{2}\frac{\partial}{\partial\mu}\left[(1-\mu^2)\frac{\partial F}{\partial\mu}\right].
\label{KINETIC_EQ}
\eeq

Let us consider an electron traveling in the loss cone $\mu>\mu_{\rm crit}=\sqrt{1-1/m}\approx 1-1/2m$ 
(or $\mu<-\mu_{\rm crit}$). The effective electron mean free path, which is the typical distance 
the electron travels before it is scattered by small angle collisions out of the loss cone, is 
\beq
\lambda_{\rm eff}\define\lambda/2m\ll\lambda.
\label{LAMBDA_EFF}
\eeq
In other words, $\lambda_{\rm eff}$ is a decay distance for a flow of electrons traveling in the loss cones.
The solution of equation~(\ref{KINETIC_EQ}) and, therefore, the escape time $\tau_m$, depends on the mirror 
strength $m$ and the ratio $l_m/\lambda$. There are three limiting cases for which simple approximate solutions 
exist: (1) $l_m\ll\lambda_{\rm eff}=\lambda/2m$; 
(2) $\lambda_{\rm eff}\ll l_m\ll \lambda^2/\lambda_{\rm eff}=2m\lambda$; 
and (3) $\lambda^2/\lambda_{\rm eff}\ll l_m$. 
We solve equation~(\ref{KINETIC_EQ}) for case (1) in Appendix~\ref{A_CASE_1} and for
cases (2) and (3) in Appendix~\ref{A_CASES_2_3}, and we obtain the electron escape times
\beq
\begin{array}{lcll}
\tau_m^{(1)}&=&\nu^{-1}\ln m,			&\quad l_m\ll\lambda_{\rm eff},\\
\tau_m^{(2)}&=&\nu^{-1} (l_m/\lambda_{\rm eff})=\nu^{-1} (2ml_m/\lambda), 
						&\quad \lambda_{\rm eff}\ll l_m\ll\lambda^2/\lambda_{\rm eff},\\
\tau_m^{(3)}&=&\nu^{-1} (3/\pi^2){(l_m/\lambda)}^2,	&\quad \lambda^2/\lambda_{\rm eff}\ll l_m.
\end{array}
\label{TAU_CASE}
\eeq
The following simple physical arguments help to understand these results in these three limiting cases. The 
collisional scattering is
a two-dimensional random walk of a unit vector (which is the direction of the electron velocity) 
on a surface of a unit-radius sphere with frequency $\nu$ (so, the scattered angle 
$\Delta_{\rm s}=\sqrt{2\nu t}$ after time interval $t$). The right hand side of the kinetic
equation~(\ref{FULL_KINETIC_EQ}) represents a one-dimensional random walk in $\mu$-space that follows from 
the two-dimensional walk because of symmetry. However, it is convenient for the moment to return to the 
original two-dimensional scattering because it is isotropic. The angular sizes of the two loss cones on 
the unit-radius sphere are $\Delta_{\rm esc}\approx 1/\sqrt{m}$. 
First, in the limit $l_m\ll\lambda_{\rm eff}$, collisions are very weak, and the scattered angle over the 
travel time between mirrors, $l_m/V$, is $\sim \sqrt{l_m/\lambda}\ll \Delta_{\rm esc}$. Therefore, in this 
case we can disregard the electron motion in $x$-space. We divide the surface of the unit-radius sphere 
into $\sim m$ boxes, each of angular size $\sim\Delta_{\rm esc}\approx 1/\sqrt{m}$. The time it takes 
for the unit vector to random walk from one box to another is $\sim \nu^{-1}/m$, resulting in the total 
escape time $\tau_m\sim m\times(\nu^{-1}/m)=\nu^{-1}$. Because the unit vector can ``visit'' each box more 
than once, the exact result contains the logarithm of $m$.
Second, in the limit $\lambda_{\rm eff}\ll l_m\ll\lambda^2/\lambda_{\rm eff}$, we have to consider motion 
in $x$-space as well. In this case the electrons move in three-dimensional phase space, and they escape 
when they are in the two loss cones within distance $\lambda_{\rm eff}$ from the mirrors, as shown by the 
shaded regions in Figure~\ref{FIG_MIRRORS}(b). We divide the three-dimensional phase space into 
$\sim (l_m/\lambda_{\rm eff})(1/\Delta_{\rm esc}^2)\sim m^2 l_m/\lambda$ boxes, each of size 
$\lambda_{\rm eff}\Delta_{\rm esc}^2\sim \lambda/m^2$. The time it takes to move from one box to another is 
$\sim \nu^{-1}/m$, resulting in the total escape time 
$\tau_m\sim (m^2 l_m/\lambda)\times(\nu^{-1}/m)=\nu^{-1}(ml_m/\lambda)$. Note, that the electron distribution 
function is almost constant in the phase space in this case (see Appendix~\ref{A_CASES_2_3}).
Third, in the limit $\lambda^2/\lambda_{\rm eff}\ll l_m$, the escape of electrons is controlled by slow 
diffusion in $x$-space, so the escape time is approximately equal to the time of diffusion between mirrors, 
$\tau_m\sim \nu^{-1} {(l_m/\lambda)}^2$ in this case.

\begin{figure}[t]
\vspace{7.0cm}
\includegraphics{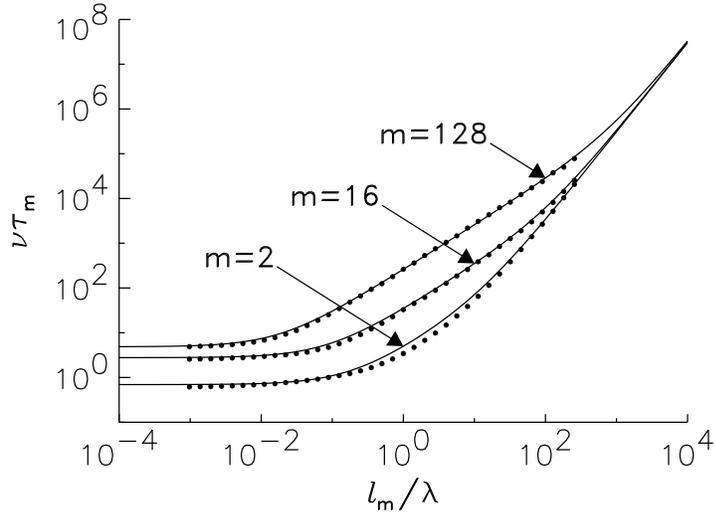}
\caption{The dots show a logarithmic plot of the numerically obtained electron escape time $\tau_m$ in 
units of the collision time $\nu^{-1}$ as a function of the separation $l_m$ of two equal magnetic mirrors 
in units of the mean free path $\lambda$. 
These results are based on our Monte-Carlo particle simulations of {$10^3$--$10^6$} trapped electrons, 
assuming three values of the mirror strengths, $m=2$, $m=16$ and $m=128$. The solid lines represent the 
analytical result, equation~(\ref{TAU}).}
\label{FIG_TAU}
\end{figure}

In our further calculations we use a simple interpolation formula
\beq
\tau_m\approx \tau_m^{(1)}+\tau_m^{(2)}+\tau_m^{(3)}=
\nu^{-1}\left[\,\ln m+(l_m/\lambda_{\rm eff})+(3/\pi^2){(l_m/\lambda)}^2\,\right]
\label{TAU}
\eeq
for the whole range of parameters $m$ and $l_m/\lambda$. This formula is suggested by the
numerical simulations shown in Figure~\ref{FIG_TAU}. The dots in this figure show the results of our 
Monte-Carlo particle simulations for three mirror strengths $m=2$, $m=16$ and $m=128$. To obtain these 
results we followed  {$10^3$--$10^6$} electrons trapped between two equal magnetic mirrors separated by 
distance $l_m$ ranging from $1/1024$ to $256$ in units of the mean free path $\lambda$. Independently of 
the initial distribution of electrons, the number of trapped electrons tends to an exponential dependence 
on time with the characteristic decay time $\tau_m$ in just a few collision times [see 
equation~(\ref{F_TIME_DEPENDENT})]. The solid lines in the figure represent formula~(\ref{TAU}) and
are in a very good agreement with the simulations even for the smallest mirror strength $m=2$.


\section{Diffusion of mono-energetic electrons in a system of random magnetic mirrors}\label{DIFFUSION}

In this section we continue to assume that electrons have a single value of speed, $V$.
If there were no magnetic mirrors and the magnetic field had constant strength along the field lines, 
the parallel diffusion of mono-energetic electrons would be the standard spatial diffusion, 
$D_0=(1/3)V\lambda$. Here, $\lambda$ is the electron mean free path at speed $V$.
However, as we have discussed in the introductory section, diffusing electrons move along flux tubes 
with random magnetic field strength and become trapped and detrapped between magnetic mirrors. These mirrors 
are regions of strong field and are separated by a field decorrelation length $l_0$. As a result, the diffusion 
is reduced by a factor that depends on the ratio $l_0/\lambda$.

In the main part of this section we derive this diffusion analytically and at the end of the section
confirm it with numerical simulations. (In contrast to the previous section, where there were only two
equal mirrors, in this section, we consider many mirrors with random spacing and strength.)

Consider the limit $l_0\gg\lambda$ first. In this case collisions are strong, and according to 
the third formula in equation~(\ref{TAU_CASE}), the time it takes for electrons to escape a trap between
two magnetic mirrors is independent of the mirror strengths and is entirely controlled by the standard 
spatial diffusion transport of electrons between the mirrors. As a result, magnetic mirrors can be ignored, 
and there is no reduction of diffusion; $D=D_0$.

In the opposite limit, $l_0\ll\lambda$, the collisions are weak, and magnetic mirrors do result in a 
reduction of diffusion. To find this reduction, we divide all mirrors into equal size bins 
$b_m=(m-\delta/2, m+\delta/2\,]$, where $m$ is the bin central mirror strength, and constant $\delta$ 
is the width of the bins (the value of $\delta$ will be discussed later). 

For the moment we consider the diffusion in the presence of only those mirrors that are in a single bin
$b_m$. It turns out that one of the bins leads to a smaller diffusion than any other bin, and the net 
diffusion due to all the mirrors is approximately that due to only mirrors in this bin, provided that the 
bins are sufficiently wide.

Let the spectrum of magnetic mirror strengths be ${\cal P}(m)$. We assume that strong magnetic mirrors 
are rare, i.e.~the spectrum falls off fast with the mirror strength (we will estimate how fast it should 
fall off, below). The probability that a mirror belongs to bin $b_m$ is
\beq
p_m = \int_{m-\delta/2}^{m+\delta/2} {\cal P}(m')\,dm'\approx \delta\,{\cal P}(m)+
(\delta^3/24){\cal P}''(m).
\label{P_M}
\eeq
At each decorrelation length $l_0$ the magnetic field changes and becomes decorrelated. Therefore, the mean 
separation of mirrors that are in bin $b_m$ is
\beq
l_m = l_0\sum_{k=1}^\infty k p_m {(1-p_m)}^{k-1}=l_0/p_m.
\label{L_M}
\eeq

Let us consider an electron trapped between two mirrors of bin $b_m$. The time $\tau_m$ that it takes for 
this electron to escape the trap is given by equation~(\ref{TAU}), where we keep only the first two 
terms (because $l_0\ll\lambda$)
\beq
\tau_m\approx\tau_m^{(1)}+\tau_m^{(2)}=\nu^{-1}\ln{(m q_m)}.
\label{TIME_OF_DETRAP}
\eeq
Here, we introduce the important parameter
\beq
q_m\define \exp{(l_m/\lambda_{\rm eff})}=\exp{(2ml_0/p_m\lambda)},
\label{Q_M}
\eeq
where the mean distance $l_m$ between the two mirrors is given by equation~(\ref{L_M}). 
After the electron escapes, it travels freely in the loss cone in one of the two directions along the 
magnetic field lines until it is again trapped between another two mirrors of bin~$b_m$. The freely 
traveling electron becomes first trapped with probabilities $1-e^{-l_m/\lambda_{\rm eff}}=1-q_m^{-1}$ 
in $0\le x<l_m$, $e^{-l_m/\lambda_{\rm eff}}-e^{-2l_m/\lambda_{\rm eff}}=q_m^{-1}-q_m^{-2}$ in
$l_m\le x<2l_m$, $e^{-2l_m/\lambda_{\rm eff}}-e^{-3l_m/\lambda_{\rm eff}}=q_m^{-2}-q_m^{-3}$ in
$2l_m\le x<3l_m$, and so on. Therefore, the mean distance squared ${\langle{\Delta x}^2\rangle}_m$ that the 
electron travels in the loss cones before trapping is
\beq
{\langle{\Delta x}^2\rangle}_m\approx l_m^2\sum_{k=1}^\infty k^2 (q_m^{-k+1}-q_m^{-k})=
					l_m^2\,\frac{q_m(q_m+1)}{(q_m-1)^2}.
\label{STEP_SQUARED}
\eeq
The processes of trapping and detrapping repeat in time intervals $\tau_m$.
In other words, electrons random walk along the field lines in a system of mirrors that belong to bin 
$b_m$ with steps $\approx{\langle{\Delta x}^2\rangle}_m$ in time intervals $\approx\tau_m$. As a 
result, the diffusion coefficient for these electrons is $D(m)=C\,[{\langle{\Delta x}^2\rangle}_m/2\tau_m]$, 
where we introduce a scaling constant $C$, which is of the order unity and will be determined by the 
numerical simulations. The corresponding reduction of diffusion is
\beq
D(m)/D_0=C\,\frac{3}{2}\,{\left(\frac{l_0}{\lambda}\right)}^2\,\frac{q_m(q_m+1)}{(q_m-1)^2}\,
			\frac{1}{p_m^2}\,\frac{1}{\ln{(m q_m)}}, \quad l_0\ll\lambda,
\label{D_M/D_0}
\eeq
where we use $D_0=(1/3)\nu\lambda^2$ and equations~(\ref{L_M}),~(\ref{TIME_OF_DETRAP}) 
and~(\ref{STEP_SQUARED}); $p_m$ and $q_m$ are given by equations~(\ref{P_M}) and~(\ref{Q_M}). 

For a given spectrum of mirrors ${\cal P}(m)$ and given constants $\delta$ and $C$, the diffusion 
reduction~(\ref{D_M/D_0}) due to mirrors of bin $b_m$, is a function of mirror strength $m$. Let us
analyze this function in two limits: $\ln q_m\ll 1$ and $\ln q_m\gg \ln m\simgt 1$.
If $\ln q_m\ll 1$, then $q_m-1=2ml_0/p_m\lambda\ll 1$. Therefore, 
$D(m)/D_0\approx C(3/4)(1/m^2\ln m)$ and $(d/dm)[D(m)/D_0]<0$. 
On the other hand, if $\ln q_m\gg \ln m$, then $D(m)/D_0\approx C(3/4)(l_0/\lambda)(1/m p_m)$. Therefore, 
$(d/dm)[D(m)/D_0]>0$ if the spectrum of mirrors falls off faster than $1/m$ with the mirror 
strength.\footnote{This criterion is different from the result of Albright {\it et al.}~(2000), who 
found $1/m^2$ to be the boundary spectrum for the transition between their diffusive and subdiffusion 
regimes. We believe that the difference arises because, for flat spectra, our bin width $\delta$ starts 
to depend on $l_0/\lambda$ (and our simple diffusion model breaks down).} 
In this paper we make an assumption that the spectrum falls off significantly faster than $1/m$.

Therefore, a minimum of $D(m)/D_0$ exists. Let this minimum be achieved at $m=m_p$. Then
$\ln q_{m_p}=l_{m_p}/\lambda_{\rm eff}\sim 2/\ln m_p\sim 1$, or $l_{m_p}\sim \lambda_{\rm eff}$.
The minimum can roughly be estimated as $D(m_p)/D_0=\min{\{D(m)/D_0\}}\sim 1/m_p^2$, which is 
in agreement with the qualitative results of Albright {\it et al.}~(2000).

\begin{figure}[p]
\vspace{6.7cm}
\includegraphics{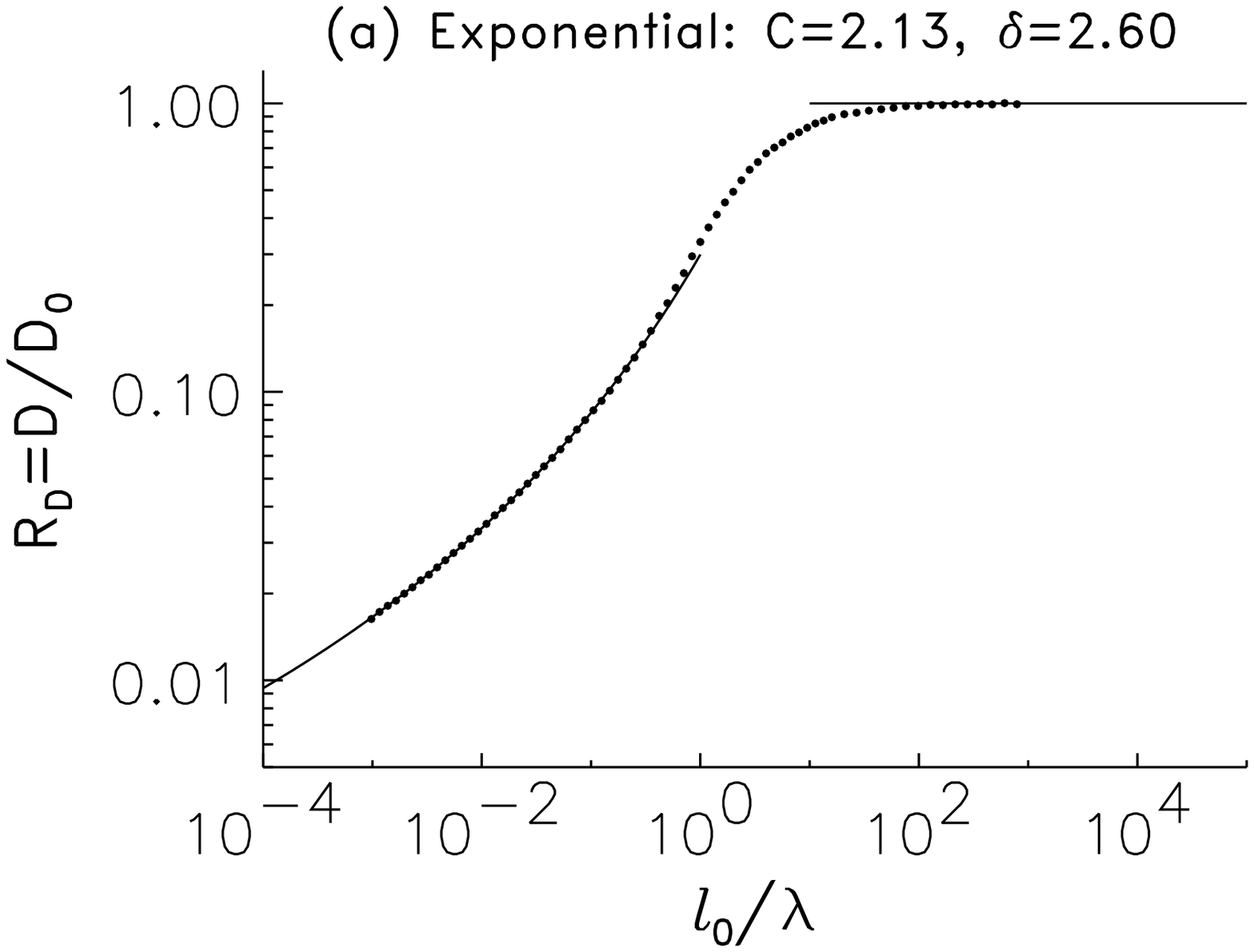}
\includegraphics{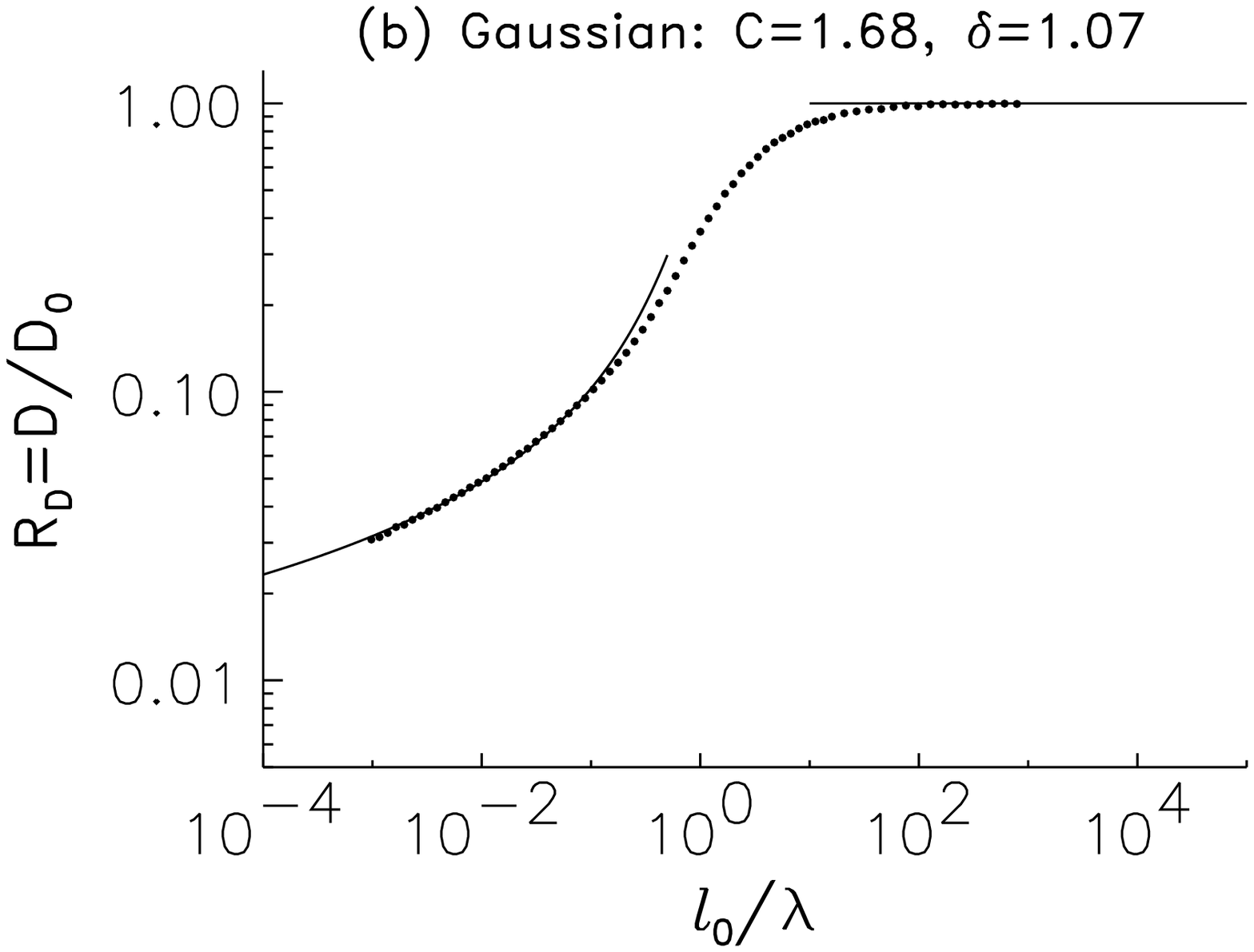}
\caption{We consider two mirror spectra: (a) exponential, and (b) Gaussian [see eqs.~(\ref{SPECTRA})]. 
The dots show the reduction of diffusion, $R_{\rm D}=D/D_0$,
obtained by Monte-Carlo particle simulations of {$1$--$6\times10^5$} electrons, each followed in a system 
of magnetic mirrors over $300$ collision times $\nu^{-1}$. The solid lines represent the theoretical 
results given by equation~(\ref{D/D_0}). The constants $C$ and $\delta$ are obtained by matching the 
theoretical results with the results of simulations for each of the two spectra (and these constants
do not depend on $l_0/\lambda$).}
\label{FIG_D/D_0}
\end{figure}

In other words, if $l_0\ll\lambda$, then there is the bin that inhibits diffusion the most. We call it 
the principle bin, $b_{\rm p}=(m_{\rm p}-\delta/2, m_{\rm p}+\delta/2\,]$. The corresponding mirror 
strength $m_{\rm p}$ is the principle mirror strength. The minimum of diffusion $D(m)$ 
due to mirrors of bin $b_m$ is achieved at the principle strength, $m=m_p$. The spacing of mirrors that 
are in the principle bin is of the order of the effective mean free path for this bin, 
$l_{m_p}\sim\lambda_{\rm eff}=\lambda/2m_p$. 
The main idea is that, in order to estimate the net diffusion due to all mirrors, we need consider only 
magnetic mirrors that are in the principle bin and we can neglect all other bins. 
Mirrors that are smaller than the principle mirrors ``work'' poorly in the inhibition of diffusion because 
they are weak and are separated by distances less than $\lambda_{\rm eff}$ (which is the distance that
electrons travel in the loss cones). Mirrors that are larger than the principle mirrors 
``work'' poorly, because they are very rare and are separated by very large distances (provided the 
mirror spectrum falls off with the mirror strength significantly faster than $1/m$). These assumptions are 
supported by our numerical simulations (see Figure~\ref{FIG_D/D_0}).

As a result of these considerations, we can combine our theoretical results for the reduction of diffusion 
of mono-energetic electrons, $R_{\rm D}=D/D_0$, into a single formula valid in the two limits for 
$l_0/\lambda$:
\beq
R_{\rm D}=D/D_0=\left\{
\begin{array}{ll}
\min\limits_m{\{D(m)/D_0\}}=D(m_p)/D_0,	&\quad l_0\ll\lambda,\\
1, 							&\quad l_0\gg\lambda,
\end{array}
\right.
\label{D/D_0}
\eeq
where $D(m)/D_0$ is given by equation~(\ref{D_M/D_0}), and the minimum is achieved at the principle 
mirror strength $m=m_p$ (note that $\ln q_{m_p}=l_{m_p}/\lambda_{\rm eff}\sim 1$).

\begin{figure}[p]
\vspace{6.7cm}
\includegraphics{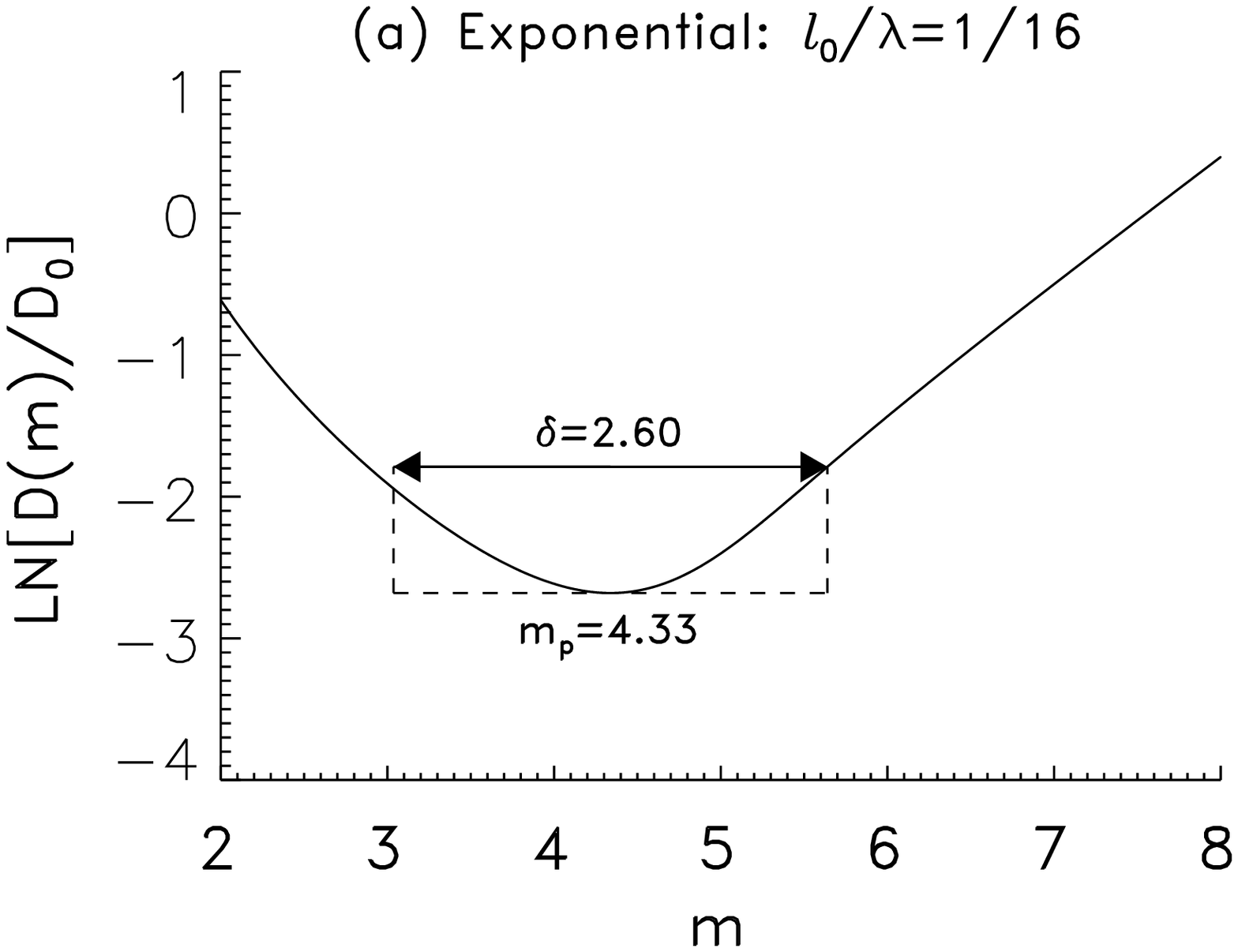}
\includegraphics{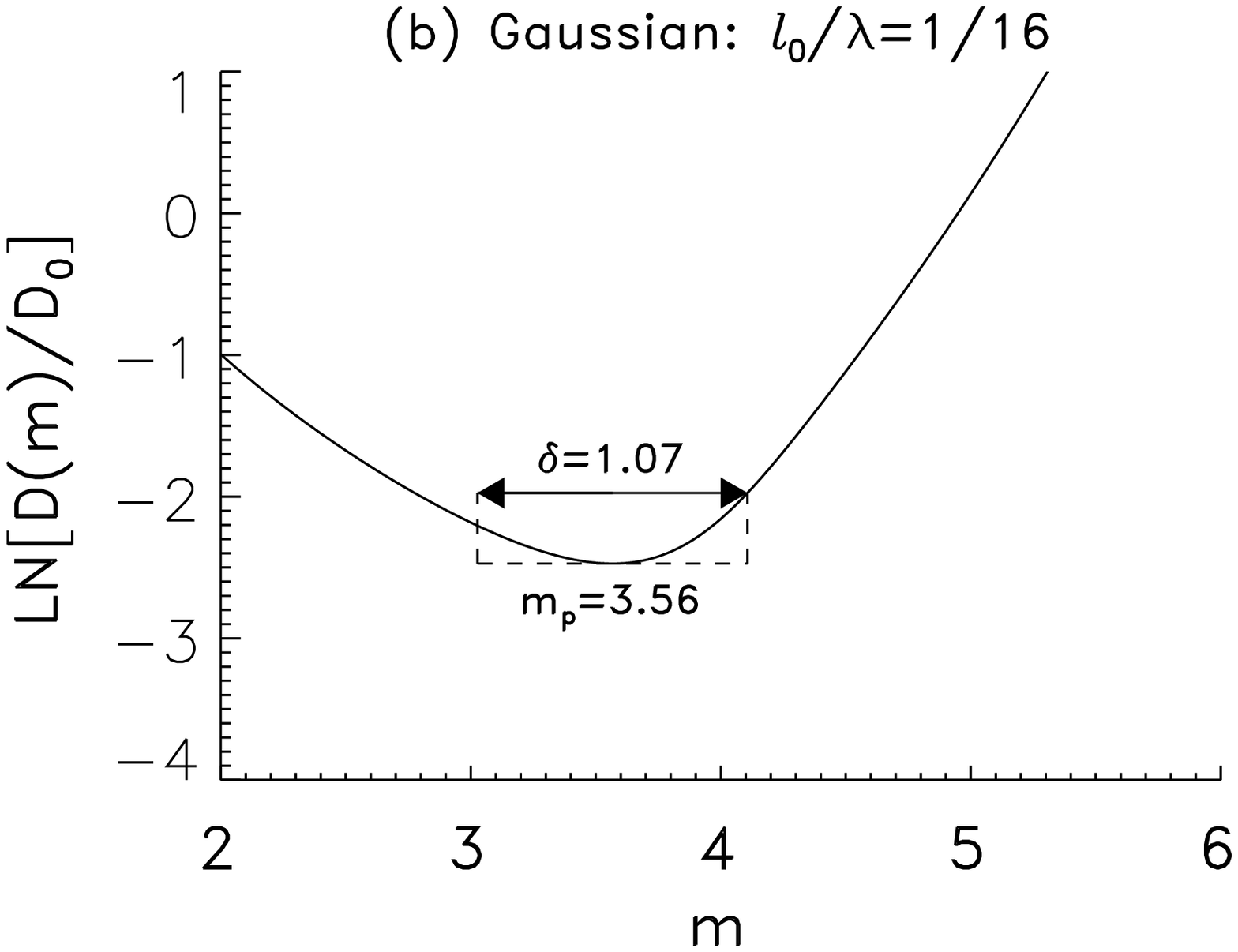}
\caption{The natural logarithm of the diffusion reduction~(\ref{D_M/D_0}) caused by mirrors that 
are in bin $b_m$ for $l_0/\lambda=1/16$. We consider two mirror spectra: 
(a) exponential, and (b) Gaussian [see eqs.~(\ref{SPECTRA})].
The principle bins are shown by arrows. In case of each spectrum, the reduction has the minimum at 
the principle mirror strength $m_p$, and it roughly doubles at the boundaries of the principle bin.}
\label{FIG_D_M/D_0}
\end{figure}

We show the theoretical mono-energetic diffusion reduction~(\ref{D/D_0}) by the solid lines in 
Figure~\ref{FIG_D/D_0} for two mirror spectra: exponential and Gaussian\footnote{We find the minimum in 
equation~(\ref{D/D_0}) numerically.},
\beq
\begin{array}{lcl}
{\cal P}(m)&=&e^{-(m-2)} \;\mbox{ exponential,}\\
{\cal P}(m)&=&{(2/\pi)}^{1/2}\,e^{-(m-2)^2/2} \;\mbox{ Gaussian}.
\end{array}
\label{SPECTRA}
\eeq
The results of our Monte-Carlo particle simulations are shown by dots. 
The constants $C$ and $\delta$ (shown at the top) are of the order unity, and we adjust them by matching our 
theoretical results with the results of simulations in case of each of the two spectra ($C$ and $\delta$ do not 
depend on $l_0/\lambda$). The simulations are based on {$1$--$6\times10^5$} particles. For each particle we 
choose a distribution of mirrors $m\ge 2$, which all are separated by the magnetic field decorrelation 
length $l_0$, and are chosen according to the assumed mirror spectrum~(\ref{SPECTRA}). We follow the particles 
during $300$ collision times $\nu^{-1}$. Then we average the particle displacements squared 
$\langle{\Delta x}^2\rangle$ at a given time $t$ to obtain the diffusion coefficient 
$\langle{\Delta x}^2\rangle/2t$ given in Figure~\ref{FIG_D/D_0}.

Note that the bin width $\delta$ is larger for the exponential spectrum than it is for the Gaussian. This 
is because the later is steeper at large mirror strengths. Figures~\ref{FIG_D_M/D_0}(a) and~\ref{FIG_D_M/D_0}(b)
clearly demonstrate the difference. In these figures we plot the natural logarithm of the diffusion 
reduction~(\ref{D_M/D_0}) caused by mirrors that are in bin $b_m$ versus the mirror strength $m$ for
$l_0/\lambda=1/16$ and for the both spectra~(\ref{SPECTRA}) of mirror strengths. The principle bins are shown 
by arrows. In case of each spectrum, the reduction has the minimum at the corresponding principle mirror 
strength $m_p$. We see, that the reduction roughly doubles over its minimal value at the
boundaries of the principle bin, $m=m_p+\delta/2$ and $m=m_p-\delta/2$.


\section{\bf The Fokker-Planck kinetic equation}\label{EQUATION}

In this section we use the results found above to obtain the modified kinetic equation for 
electrons traveling in a system of random magnetic mirrors. The reduction of diffusion of mono-energetic 
electrons with speed $V$, $R_{\rm D}$, obtained in the previous section can be considered to be 
equivalent to an enhancement of the pitch angle scattering rate, since the pitch angle scattering is
directly related to spatial diffusion. We therefore, in deriving the collision integral, modify the pitch
angle scattering term by factor $R_{\rm D}^{-1}$, where $R_{\rm D}$ is the factor by which the spatial 
diffusion is reduced (see the previous section). Hereafter, we do not assume electrons to be 
mono-energetic. We take into account the full perturbed electron-electron collision integral, as well 
as the electron-proton collision term. When $R_{\rm D}\equiv 1$, our equations reduce to those of Spitzer 
and H$\ddot{\rm a}$rm in their well known paper 
(Spitzer \& H$\ddot{\rm a}$rm 1953; Cohen, Spitzer \& Routly 1950; Spitzer 1962).

The electron distribution function is 
\beq
f(\mu,V)=f_0(V)+f_1(\mu,V),
\label{DISTRIBUTION_FUNCTION}
\eeq
where $f_0$ is the zero order isotropic part given by the Maxwellian distribution,
\beq
f_0=n(x)\,{\left[m_e/2\pi kT(x)\right]}^{3/2}\,e^{-m_eV^2/2kT(x)}=
n\pi^{-3/2}V_{\rm T}^{-3}\,e^{-\upsilon^2},
\label{MAXWELL}
\eeq
and $f_1\propto\mu$ is the first order anisotropic perturbation (of order the temperature gradient and
electric field)
\beq
f_1(\mu,V)=\mu n V_{\rm T}^{-3} S(\upsilon).
\label{F_1}
\eeq
Here $m_e$ is the electron mass, $k$ is the Boltzmann constant, and the electron temperature $T(x)$ and 
concentration $n(x)$ slowly change in space. We also introduce the dimensionless electron speed 
$\upsilon=V/V_T$, where the thermal electron speed is $V_T=\sqrt{2kT/m_e}$. Thus, the function $S(\upsilon)$
in equation~(\ref{F_1}) is dimensionless.

In a steady state, the kinetic equation for the electrons is obviously
\beq
V_x(\partial f_0/\partial x)-(eE/m_e)(\partial f_0/\partial V_x)={(\delta f/\delta t)}_c,
\label{FULL_KINETIC_EQUATION}
\eeq
where ${(\delta f/\delta t)}_c$ is the Coulomb collision integral that includes electron-proton and 
electron-electron collisions, $V_x=\mu V$ is the $x$-component of the electron velocity (the component 
along the magnetic field lines), and $E$ is the electric field in the $x$-direction. The electron pressure
should be constant, $P=k\,n(x)T(x)={\rm const}$.~\footnote{Because the hydrodynamic time scale is much 
shorter than the transport, e.g~thermal conduction, time scale.} As a result, the derivatives of the 
Maxwellian electron distribution
are
\beq
\partial f_0/\partial x=(\upsilon^2-2.5)(f_0/T)(dT/dx),
\quad
\partial f_0/\partial V_x=-(2\mu/V_{\rm T})\upsilon f_0.
\label{F_0_DERIVATIVES}
\eeq
 
The collision integral is divided up as
\beq
{(\delta f/\delta t)}_c={(\delta f_0/\delta t)}_0+{(\delta f_1/\delta t)}_0+{(\delta f_0/\delta t)}_1=
{(\delta f_1/\delta t)}_0+{(\delta f_0/\delta t)}_1,
\label{FULL_COLLISION_INTEGRAL}
\eeq
where ${(\delta f_0/\delta t)}_0\equiv 0$ corresponds to Maxwellian collisions acting on $f_0$, 
${(\delta f_1/\delta t)}_0$ corresponds to Maxwellian collisions (with enhanced pitch angle scattering) 
acting on $f_1$, and ${(\delta f_0/\delta t)}_1$ corresponds to perturbed collisions acting on $f_0$
(since $f_0$ is isotropic, there is no pitch angle scattering in this collision term).
The collision integral~(\ref{FULL_COLLISION_INTEGRAL}) can be best obtained, in the Fokker-Planck form, 
by using the Rosenbluth potentials $h(\mu,V)=h_0(V)+h_1(\mu,V)$ and $g(\mu,V)=g_0(V)+g_1(\mu,V)$ 
(Rosenbluth, MacDonald, \& Judd 1957). Here $h_0$ and $g_0$ are calculated using the Maxwellian 
parts of the electron and ion distribution functions~(\ref{MAXWELL}), while the perturbed potentials,  
$h_1=2\mu{\cal A}_1(V)$ and $g_1=\mu{\cal B}_1(V)$, are proportional to $\mu$, and they are calculated 
using the perturbed part of the electron distribution function~(\ref{F_1}). 

The Maxwellian potentials $h_0$ and $g_0$ determine the ${(\delta f_1/\delta t)}_0$ part of the Fokker-Planck 
collision integral, and the perturbed potentials, $h_1=2\mu{\cal A}_1(V)$ and $g_1=\mu{\cal B}_1(V)$, are 
used to find the ${(\delta f_0/\delta t)}_1$ part of the Fokker-Planck collision integral 
[see the equation~(31) of Rosenbluth, MacDonald, \& Judd 1957]
\beq
{(\delta f_1/\delta t)}_0&=&\frac{A_D}{2n}\left\{
-\frac{1}{V^2}\frac{\partial}{\partial V}\left[f_1V^2\frac{dh_0}{dV}\right]
+\frac{1}{2V^2}\frac{\partial^2}{\partial V^2}\left[f_1V^2\frac{d^2g_0}{dV^2}\right]
\right.\nonumber\\
&-&\left.
\frac{1}{V^2}\frac{\partial}{\partial V}\left[f_1\frac{dg_0}{dV}\right]
+R_{\rm D}^{-1}\frac{1}{2V^3}\frac{dg_0}{dV}\frac{\partial}{\partial\mu}
\left[(1-\mu^2)\frac{\partial f_1}{\partial\mu}\right]
\right\},
\label{FULL_COLL_INT_1_0}
\\
{(\delta f_0/\delta t)}_1&=&\mu\frac{A_D}{2n}\left\{
-\frac{2}{V^2}\frac{d}{dV}\left[f_0V^2\frac{d{\cal A}_1}{dV}\right]
+\frac{4}{V^2}f_0{\cal A}_1
+\frac{1}{2V^2}\frac{d^2}{dV^2}\left[f_0V^2\frac{d^2{\cal B}_1}{dV^2}\right]
\right.\nonumber\\
&-&\left.
\frac{3}{V^3}f_0\frac{d{\cal B}_1}{dV}
+\frac{3}{V^4}f_0{\cal B}_1
-\frac{3}{V^2}\frac{d}{dV}\left[f_0\frac{d{\cal B}_1}{dV}\right]
+\frac{3}{V^2}\frac{d}{dV}\left[f_0\frac{{\cal B}_1}{V}\right]
\right\}.
\label{FULL_COLL_INT_0_1}
\eeq
For a hydrogen plasma the ``diffusion constant'' $A_D$ is
\beq
A_D=8\pi n e^4 \ln\Lambda/m_e^2,
\label{A_D}
\eeq
where $e$ is the absolute value of the electron charge, and $\ln\Lambda$ is
the Coulomb logarithm (Spitzer 1962). Note, that the last term in equation~(\ref{FULL_COLL_INT_1_0}) 
is the pitch angle scattering term, and we multiply it by our enhancement factor $R_{\rm D}^{-1}$
[compare this term with the right-hand side of equation~(\ref{FULL_KINETIC_EQ})].

Using the equations~(17) and~(18) of Rosenbluth, MacDonald, \& Judd (1957),
we express the derivatives of the potentials $h_0$ and $g_0$ in terms of the three Maxwellian 
diffusion coefficients ${\langle\Delta V_\|\rangle}_0$,
${\langle{(\Delta V_\perp)}^2\rangle}_0$ and ${\langle{(\Delta V_\|)}^2\rangle}_0$, 
which are further given in terms of error functions [see the {equations~(5-15)--(5-20)} of Spitzer 1962]
\beq
\begin{array}{lcl}
dh_0/dV&=&(2n/A_D)\,{\langle\Delta V_\|\rangle}_0=-(n/V^2)[1+4\upsilon^2G(\upsilon)],\\
dg_0/dV&=&(n/A_D)V\,{\langle{(\Delta V_\perp)}^2\rangle}_0=n[1+\Phi(\upsilon)-G(\upsilon)],\\
d^2g_0/dV^2&=&(2n/A_D)\,{\langle{(\Delta V_\|)}^2\rangle}_0=(2n/V)G(\upsilon).
\end{array}
\label{H_0_G_0}
\eeq
Here $\Phi$ is the usual error function, and $G$ is expressed in terms of $\Phi$ and its 
derivative $\Phi'$, they are functions of the dimensionless speed $\upsilon=V/V_T$ [$V_T=\sqrt{2kT/m_e}$],
\beq
&&\Phi(\upsilon)=(2/\sqrt{\pi})\int_0^\upsilon e^{-x^2}dx,
\qquad
G(\upsilon)=\frac{\Phi(\upsilon)-\upsilon \Phi'(\upsilon)}{2\upsilon^2}.
\label{PHI_G}
\eeq
The perturbed potentials, $h_1=2\mu{\cal A}_1(V)$ and $g_1=\mu{\cal B}_1(V)$, are calculated using
the perturbed electron distribution function~(\ref{F_1}) and are given by
the following formulas [see the equations~(40), (41), (45) and~(46) of Rosenbluth, MacDonald, \& Judd 1957]
\beq
{\cal A}_1&=&(4\pi/3)(n/V_{\rm T})\left[\upsilon^{-2}{\overline I}_3(S;\upsilon)
+\upsilon{\underline I}_{\,0}(S;\upsilon)\right],
\nonumber\\
{\cal B}_1&=&(4\pi/3)nV_{\rm T}\left[0.2\upsilon^{-2}{\overline I}_5(S;\upsilon)-{\overline I}_3(S;\upsilon)
-\upsilon{\underline I}_{\,2}(S;\upsilon)+0.2\upsilon^3{\underline I}_{\,0}(S;\upsilon)\right],
\label{A_1_B_1}
\eeq
where we introduce integrals
\beq
&&{\overline I}_m(S;\upsilon)=\int_0^\upsilon\!\upsilon^m S(\upsilon)\,d\upsilon,
\qquad
{\underline I}_{\,m}(S;\upsilon)=\int_\upsilon^\infty\!\upsilon^m S(\upsilon)\,d\upsilon,
\label{I_DEFINITION}
\eeq

Now, substituting equations~(\ref{MAXWELL}),~(\ref{F_1}),~(\ref{H_0_G_0}) and~(\ref{A_1_B_1}) into
formulas~(\ref{FULL_COLL_INT_1_0}) and~(\ref{FULL_COLL_INT_0_1}), and using 
definitions~(\ref{PHI_G}),~(\ref{I_DEFINITION}) and equation~(\ref{FULL_COLLISION_INTEGRAL}), after 
considerable algebra, we have for the collision integrals
\beq
{(\delta f_1/\delta t)}_0&=&(nA_D/2V_{\rm T}^6)\,\mu\upsilon^{-2}({\hat{\cal L}}S-2\upsilon^2\Phi'S),
\nonumber\\
{(\delta f_0/\delta t)}_1&=&(nA_D/2V_{\rm T}^6)\,\mu\upsilon^{-2}({\hat{\cal I}}S+2\upsilon^2\Phi'S),
\nonumber\\
{(\delta f/\delta t)}_c&=&(nA_D/2V_{\rm T}^6)\,\mu\upsilon^{-2}({\hat{\cal L}}S+{\hat{\cal I}}S),
\label{COLLISION_INTEGRAL}
\eeq
where the differential and the integral operators are defined as
\beq
{\hat{\cal L}}S(\upsilon)&=&d/d\upsilon\!\left[\upsilon G\,(dS/d\upsilon)\vphantom{\Phi'}\right]+
2\upsilon^2G\,(dS/d\upsilon)-\left[\upsilon^{-1}R_{\rm D}^{-1}(1+\Phi-G)-4\upsilon^2\Phi'\right]\!S,
\label{L_OPERATOR}\\
{\hat{\cal I}}S(\upsilon)&=&(4/15\sqrt{\pi})\,e^{-\upsilon^2}\left[12{\overline I}_5(S;\upsilon)-
10{\overline I}_3(S;\upsilon)+2\upsilon^3(6\upsilon^2-5){\underline I}_{\,0}(S;\upsilon)\right].
\label{I_OPERATOR}
\eeq
The enhancement of the Maxwellian pitch angle scattering rate, $R_{\rm D}^{-1}$, enters into the 
differential operator~(\ref{L_OPERATOR}). $R_{\rm D}$ depends on the dimensionless speed $\upsilon=V/V_T$,
we will explicitly give this dependence in equations~(\ref{R_D_LORENTZ}) and~(\ref{R_D_SPITZER}).

Finally, substituting formulas~(\ref{COLLISION_INTEGRAL}) and~(\ref{F_0_DERIVATIVES}) into
equation~(\ref{FULL_KINETIC_EQUATION}), we obtain the kinetic equation for the dimensionless perturbed 
electron distribution function $S(\upsilon)$ [see equation~(\ref{F_1})]
\beq
{\hat{\cal L}}S&=&\gamma_{\mbox{\tiny T}}\,\upsilon^3(2\upsilon^2-5)e^{-\upsilon^2}+
\gamma_{\mbox{\tiny E}}\,\upsilon^3e^{-\upsilon^2}-{\hat{\cal I}}S,
\label{KINETIC_EQUATION}\\
S(\upsilon)&\to&0, \quad\mbox{as }\upsilon\to 0\mbox{ and as }\upsilon\rightarrow \infty,
\label{BOUNDARY_CONDITIONS}
\eeq
where constants $\gamma_{\mbox{\tiny T}}$ and $\gamma_{\mbox{\tiny E}}$ are
\beq
\gamma_{\mbox{\tiny T}}=\frac{k^2T}{2\pi^{5/2}ne^4\ln\Lambda}\,\frac{dT}{dx},
\qquad
\gamma_{\mbox{\tiny E}}=\frac{kT}{\pi^{5/2}ne^3\ln\Lambda}\,E.
\label{GAMMA_T_E}
\eeq
We also take the obvious boundary conditions~(\ref{BOUNDARY_CONDITIONS}) for function $S$.
Equations~(\ref{L_OPERATOR})--(\ref{KINETIC_EQUATION}) reduce to the Spitzer equations for
an ionized hydrogen gas (Spitzer \& H$\ddot{\rm a}$rm 1953; Cohen, Spitzer \& Routly 1950) 
if we set $R_{\rm D}\equiv 1$ and make a substitution $S(\upsilon)=\pi^{-3/2}e^{-\upsilon^2}D(\upsilon)$.
However, we prefer to use function $S$, because of the simpler boundary conditions~(\ref{BOUNDARY_CONDITIONS}).


\section{\bf The reduction of transport coefficients by stochastic magnetic mirrors}\label{TRANSPORT}

In a steady state, an electric field $E$ and a temperature gradient $dT/dx$ both produce aniso\-tro\-pic 
perturbations of the electron distribution function, $f_1(\mu,\upsilon)=\mu n V_{\rm T}^{-3} S(\upsilon)$, 
see equations~(\ref{DISTRIBUTION_FUNCTION}) and~(\ref{F_1}). This anisotropy results in an electron
flow and, consequently, in an electric current $j$ and in a heat flow $Q$ along magnetic field lines 
(in the $x$-direction)
\beq
j&=&-e\int_0^\infty\!\!\int_{-1}^1 \mu V\,f_1\,d\mu\,2\pi V^2dV\,=\,\sigma E+\alpha\,(dT/dx),
\label{J}\\
Q&=&\int_0^\infty\!\!\int_{-1}^1 \mu V\,(m_eV^2/2)\,f_1\,d\mu\,2\pi V^2dV\,=\,-\beta E-\kappa\,(dT/dx).
\label{Q}
\eeq
Here $\sigma$, $\alpha$, $\beta$ and $\kappa$ are the four transport coefficients to be found 
($\sigma$ and $\kappa$ are the electrical and thermal conductivities). 

Before we proceed to the calculation of the transport coefficients, let us first call attention to
the electron flow produced by the electric field. The electric field produces two different kinds
of the electron flow. The first, the main, flow is due to acceleration of electrons, which is described 
by the term containing $E$ in equation~(\ref{FULL_KINETIC_EQUATION}), and correspondingly by the term 
containing $\gamma_{\mbox{\tiny E}}$ in equation~(\ref{KINETIC_EQUATION}). The second, an additional, 
flow arises because the electric field changes the size of the two loss cones of a mirror trap, so 
in Figure~\ref{FIG_MIRRORS}(b) $\mu_{\rm crit}$ in the right upper corner is not equal to 
$\mu_{\rm crit}$ in the left lower corner. As a result, the electrons are more likely to escape the
trap in the direction opposite to the electric field. Fortunately, this additional flow, which is rather
complicated to find precisely, can be neglected compared to the flow due to acceleration. We give a 
prove of this in Appendix~\ref{E_FLUXES}.~\footnote{The main reason is that the difference in 
the two loss cones due to electric field is inversely proportional to the electron kinetic energy, 
so the additional flow has a factor $1/V^2$ compared to a factor $1/V_{\rm T}^2$ that enters the 
main flow due to acceleration. Because both the current and the heat flow are mainly transported 
by superthermal electrons $\upsilon=V/V_{\rm T}\sim 2$, the additional flow is approximately $20\%$ of 
the main flow, see Appendix~\ref{E_FLUXES}.}

In further calculations, it is convenient to break $S(\upsilon)$ into the two separate inhomogeneous 
solutions of equation~(\ref{KINETIC_EQUATION}), which we denote as $S_T(\upsilon)$ and 
$S_E(\upsilon)$.~\footnote{The two homogeneous solutions of equation~(\ref{KINETIC_EQUATION}) must be 
excluded, because they diverge either at $\upsilon\to 0$ or at $\upsilon\to\infty$, violating the 
conditions~(\ref{BOUNDARY_CONDITIONS}), see more details in Cohen, Spitzer \& Routly 1950.}
The first solution, $S_T$, is obtained by setting $\gamma_{\mbox{\tiny T}}=1$ and 
$\gamma_{\mbox{\tiny E}}=0$, and the second solution, $S_E$, is obtained by setting 
$\gamma_{\mbox{\tiny T}}=0$ and $\gamma_{\mbox{\tiny E}}=1$, i.~e.
\beq
\begin{array}{l}
S_T(\upsilon)=S(\upsilon), \quad \mbox{when }\gamma_{\mbox{\tiny T}}=1 \mbox{ and } \gamma_{\mbox{\tiny E}}=0,
\\
S_E(\upsilon)=S(\upsilon), \quad \mbox{when }\gamma_{\mbox{\tiny T}}=0 \mbox{ and } \gamma_{\mbox{\tiny E}}=1.
\end{array}
\label{S_E_T}
\eeq
The general solution to equation~(\ref{KINETIC_EQUATION}) and the perturbed distribution
function~(\ref{F_1}) are the linear combinations of the two inhomogeneous solutions,
\beq
S(\upsilon)&=&\gamma_{\mbox{\tiny T}}S_T(\upsilon)+\gamma_{\mbox{\tiny E}}S_E(\upsilon),
\nonumber\\
f_1(\mu,\upsilon)&=&\mu n V_{\rm T}^{-3}
\left[\gamma_{\mbox{\tiny T}}S_T(\upsilon)+\gamma_{\mbox{\tiny E}}S_E(\upsilon)\vphantom{S^0}\right].
\label{F_1_GENERAL}
\eeq
In other words, $S_T$ and $S_E$ correspond to anisotropic perturbations of the electron distribution 
function, which are driven by the temperature gradient and by the electric field respectively, while
$S=\gamma_{\mbox{\tiny T}}S_T+\gamma_{\mbox{\tiny E}}S_E$ is the total anisotropic perturbation.

We now consider separately two cases: first, the Lorentz gas in a system of random mirrors, and second,
the Spitzer gas in a system of random mirrors. For the Lorentz gas, electrons are assumed only to collide 
with protons, so equations~(\ref{L_OPERATOR})--(\ref{KINETIC_EQUATION}) become greatly simplified. For the
Spitzer gas, we consider both the electron-electron the electron-proton collisions, so we solve the full 
set of our equations.


\subsection{Lorentz gas in a system of random mirrors}\label{LORENTZ_GAS}

Here we assume the electrons to collide only with protons, so we have for 
operators~(\ref{L_OPERATOR}) and~(\ref{I_OPERATOR})
\beq
{\hat{\cal L}}S=-S/\upsilon R_{\rm D}, \quad {\hat{\cal I}}S=0,
\eeq
resulting in the two simple inhomogeneous solutions~(\ref{S_E_T}) of equation~(\ref{KINETIC_EQUATION}),
\beq
S_T(\upsilon)=-\upsilon^4(2\upsilon^2-5)\,e^{-\upsilon^2}R_{\rm D}, 
\quad
S_E(\upsilon)=-\upsilon^4\,e^{-\upsilon^2}R_{\rm D}.
\label{LORENTZ_S}
\eeq
If there are no magnetic mirrors, so $R_{\rm D}\equiv 1$, we substitute equations~(\ref{LORENTZ_S}) 
into formula~(\ref{F_1_GENERAL}) and easily carry out the two integrals in equations~(\ref{J}) 
and~(\ref{Q}). Taking into consideration definitions~(\ref{GAMMA_T_E}), we obtain the well-known Lorentz 
transport coefficients (Spitzer 1962)
\beq
&&\sigma_{\mbox{\tiny L}}=2{\left(\frac{2}{\pi}\right)}^{3/2}\frac{{(kT)}^{3/2}}{m_e^{1/2}e^2\ln\Lambda},
\qquad
\alpha_{\mbox{\tiny L}}=3{\left(\frac{2}{\pi}\right)}^{3/2}\frac{{k(kT)}^{3/2}}{m_e^{1/2}e^3\ln\Lambda},
\nonumber\\
&&\beta_{\mbox{\tiny L}}=8{\left(\frac{2}{\pi}\right)}^{3/2}\frac{{(kT)}^{5/2}}{m_e^{1/2}e^3\ln\Lambda},
\qquad
\kappa_{\mbox{\tiny L}}=20{\left(\frac{2}{\pi}\right)}^{3/2}\frac{{k(kT)}^{5/2}}{m_e^{1/2}e^4\ln\Lambda}.
\label{LORENTZ_TRANSPORT}
\eeq

If there are magnetic mirrors, it is convenient to normalize the resulting transport coefficients 
to the corresponding Lorentz coefficients~(\ref{LORENTZ_TRANSPORT}). Substituting 
equation~(\ref{F_1_GENERAL}) into the two integrals in equations~(\ref{J}) and~(\ref{Q}), and 
again using definitions~(\ref{GAMMA_T_E}), we have
\beq
&&\sigma/\sigma_{\mbox{\tiny L}}=-(1/3)\,{\overline I}_3(S_E;\infty),
\qquad
\alpha/\alpha_{\mbox{\tiny L}}=-(1/9)\,{\overline I}_3(S_T;\infty),
\nonumber\\
&&\beta/\beta_{\mbox{\tiny L}}=-(1/12)\,{\overline I}_5(S_E;\infty),
\qquad
\kappa/\kappa_{\mbox{\tiny L}}=-(1/60)\,{\overline I}_5(S_T;\infty),
\label{TRANSPORT_COEFFICIENTS}
\eeq
where the integral moments are defined by equations~(\ref{I_DEFINITION}), and $S_T$ and $S_E$ are given 
by equations~(\ref{LORENTZ_S}).

In order to find explicitly the diffusion reduction factor $R_{\rm D}$ in equations~(\ref{LORENTZ_S}) 
as a function of $\upsilon$, we refer to the results of Section~\ref{DIFFUSION}. In those section we
found the diffusion reduction as a function of the ratio of the magnetic field decorrelation length
$l_0$ to the electron mean free path $\lambda$. For Lorentz electrons the mean free path $\lambda_{\rm L}$
is proportional to the fourth power of the electron speed, $\lambda_{\rm L}\propto V^4$, (Spitzer 1962, 
Braginskii 1965). Thus, we have 
\beq
R_{\rm D}=R_{\rm D}(l_0/\lambda_{\rm L})=R_{\rm D}(\upsilon^{-4}l_0/\lambda_{{\rm L},T}),
\label{R_D_LORENTZ}
\eeq
where $\lambda_{{\rm L},T}$ is obviously the Lorentz electron mean free path at the thermal speed 
$V_T=\sqrt{2kT/m_e}$
\beq
\lambda_{{\rm L},T}=\left.(kT)^2\right/\pi n e^4\ln\Lambda
\approx 0.1\,{\rm Kpc}\,{(T/10^7{\rm K})}^2(10^{-3}{\rm cm}^{-3}/n).
\label{LAMBDA_L_T}
\eeq
Here we assume the Coulomb logarithm for a cluster of galaxy to be $\ln\Lambda\approx 40$
(Suginohara \& Ostriker 1998).

We use our theoretical results given by equation~(\ref{D/D_0}) for the mono-energetic diffusion reduction 
$R_{\rm D}=R_{\rm D}(l_0/\lambda_{\rm L})=R_{\rm D}(\upsilon^{-4}l_0/\lambda_{{\rm L},T})$ in the limits 
$\upsilon^{-4}l_0/\lambda_{{\rm L},T}\ll 1$ and $\upsilon^{-4}l_0/\lambda_{{\rm L},T}\gg 1$; 
and we use our numerical simulation results presented in Figure~\ref{FIG_D/D_0} for 
$\upsilon^{-4}l_0/\lambda_{{\rm L},T}\sim 1$. [We carry out the cubic spline interpolation of the 
simulation results. Note, that $R_{\rm D}$ is not differentiated in operator~(\ref{L_OPERATOR}), so
our final results are not sensitive to small noise errors in calculation of $R_{\rm D}$.]

\begin{figure}[p]
\vspace{17.2cm}
\includegraphics{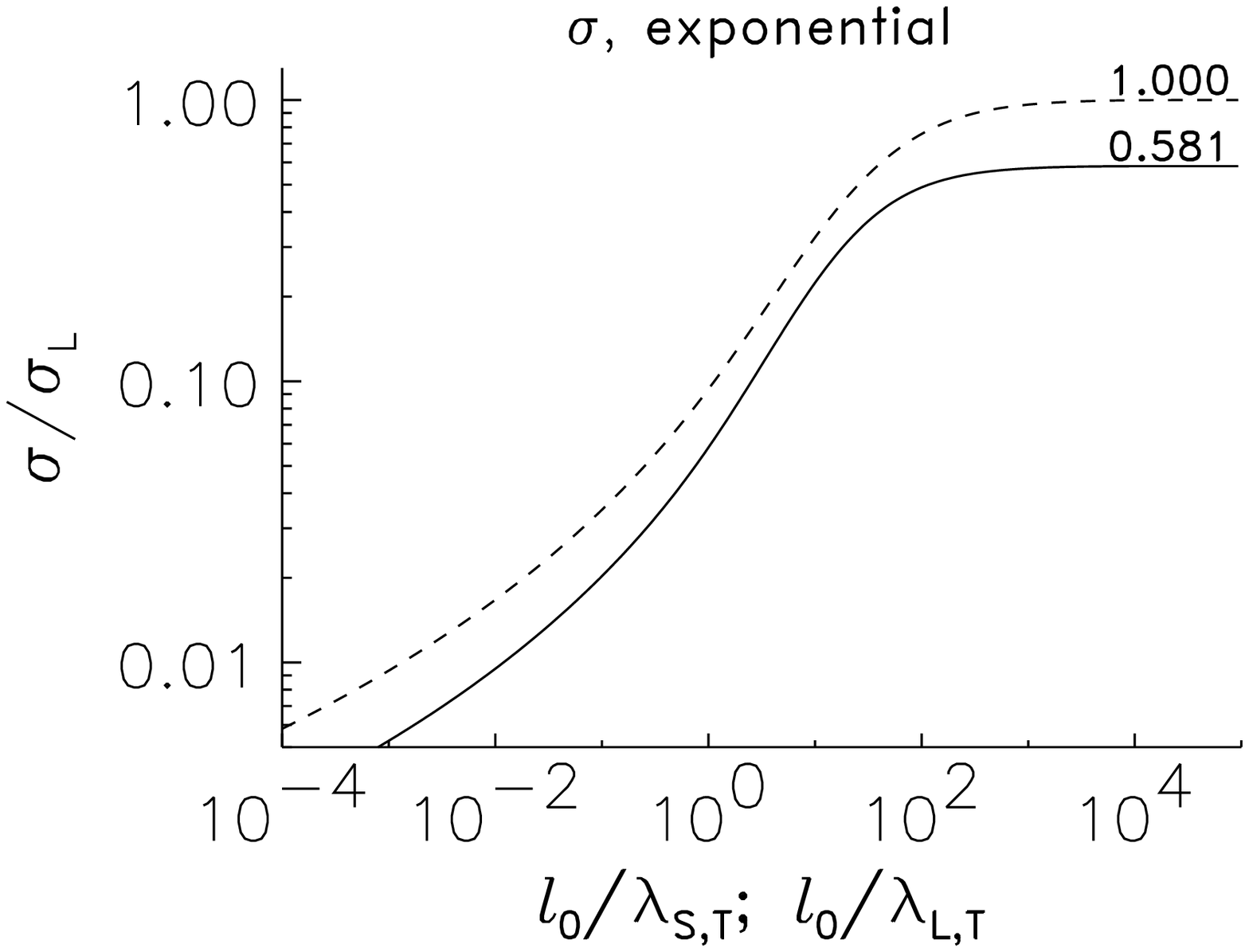}
\includegraphics{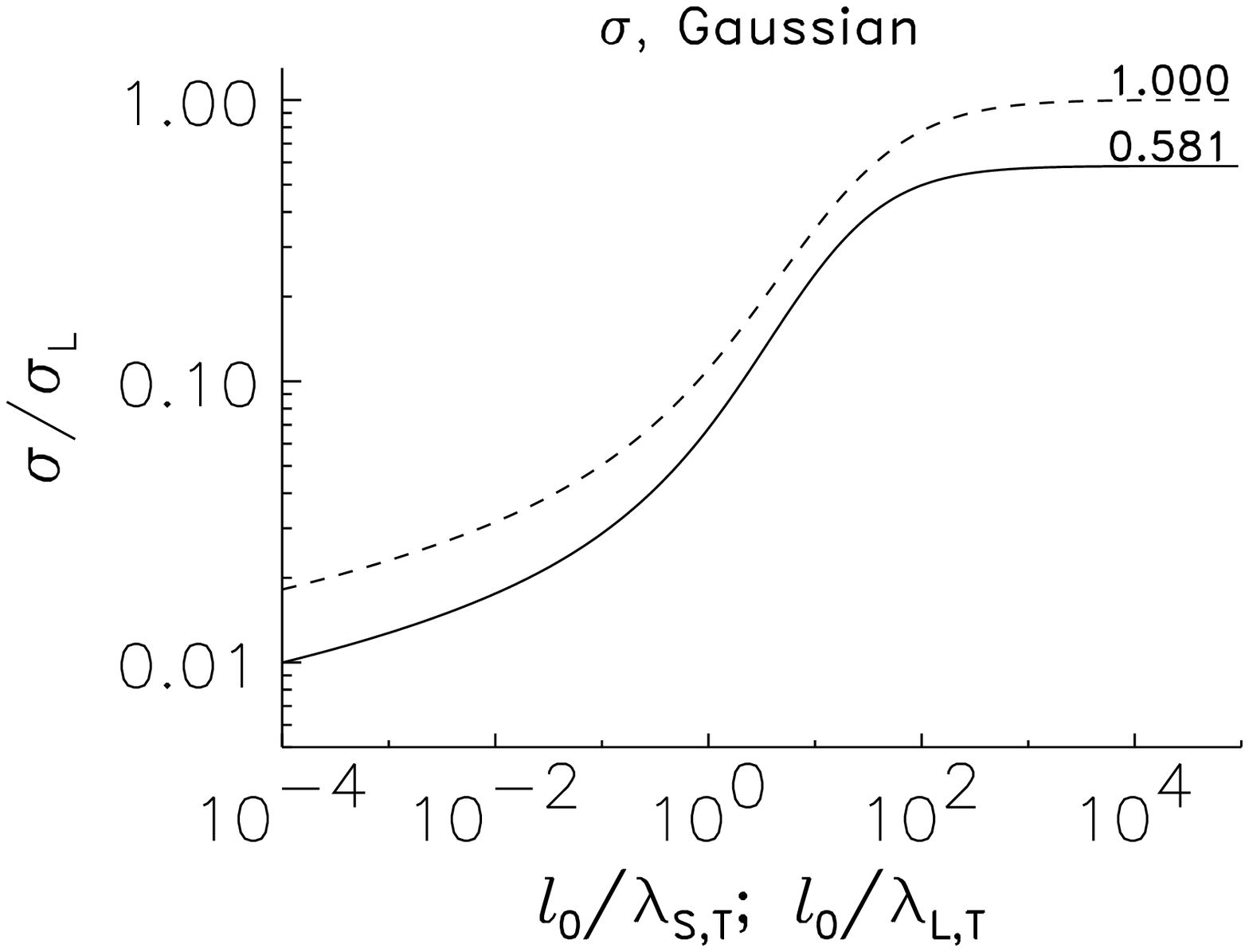}
\includegraphics{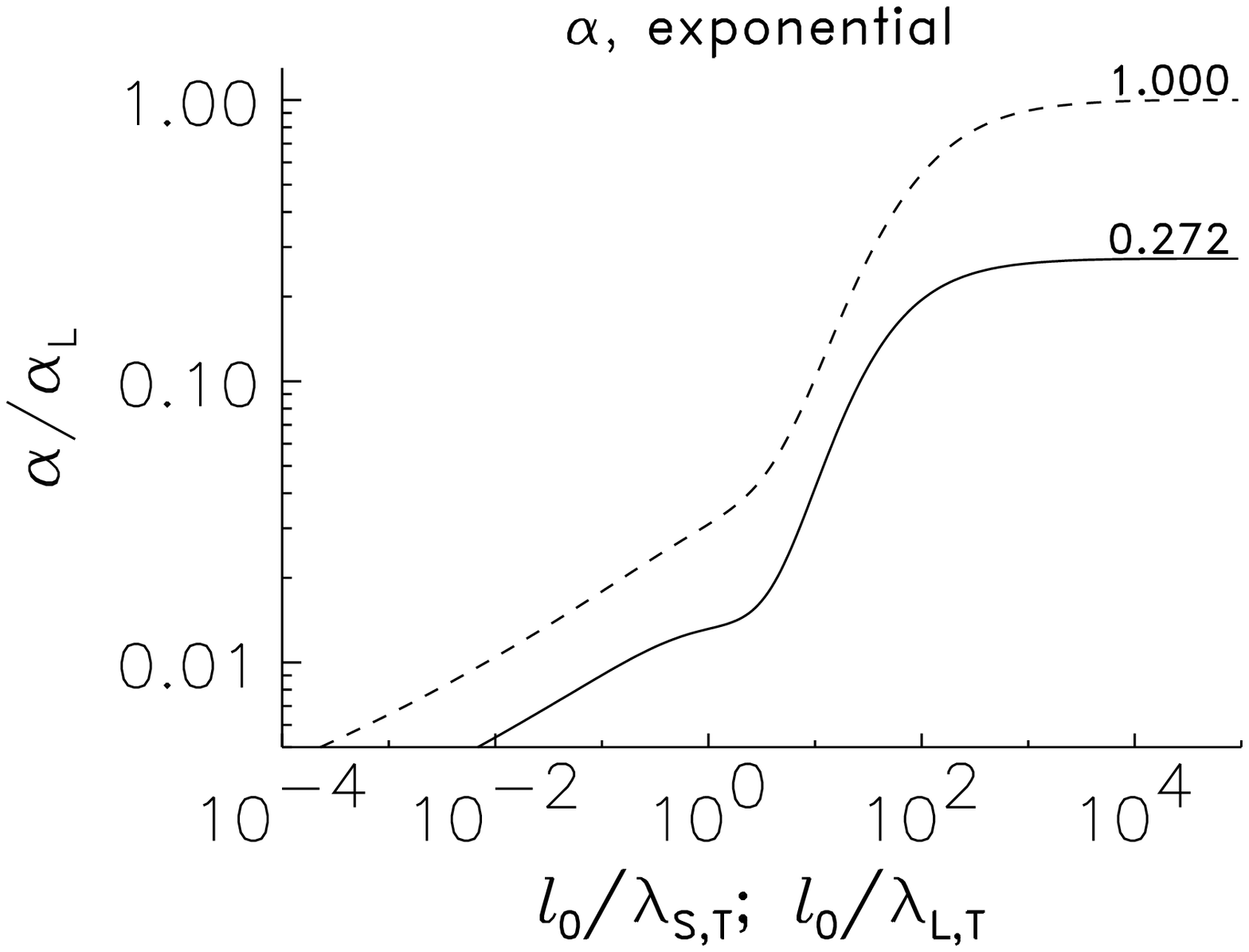}
\includegraphics{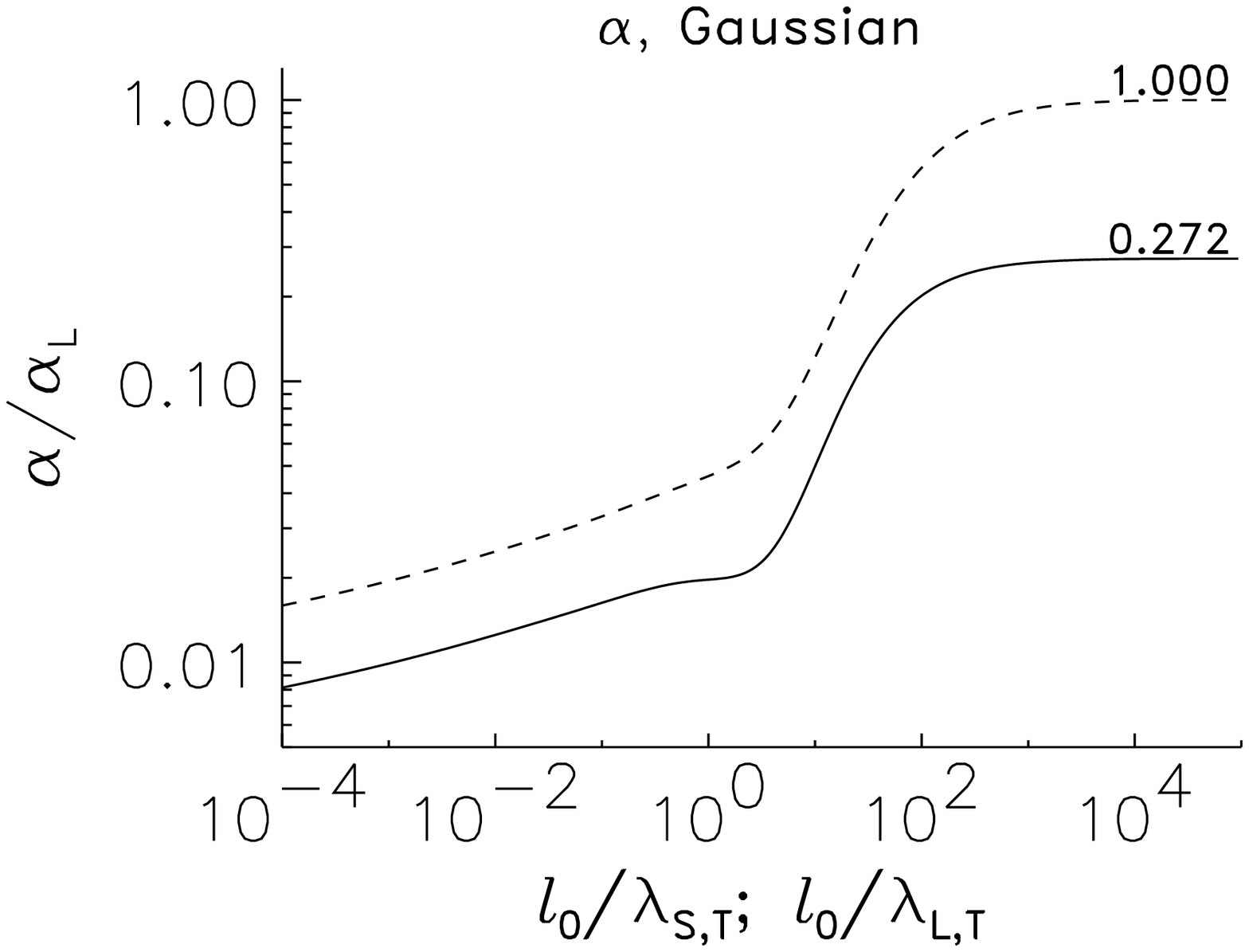}
\includegraphics{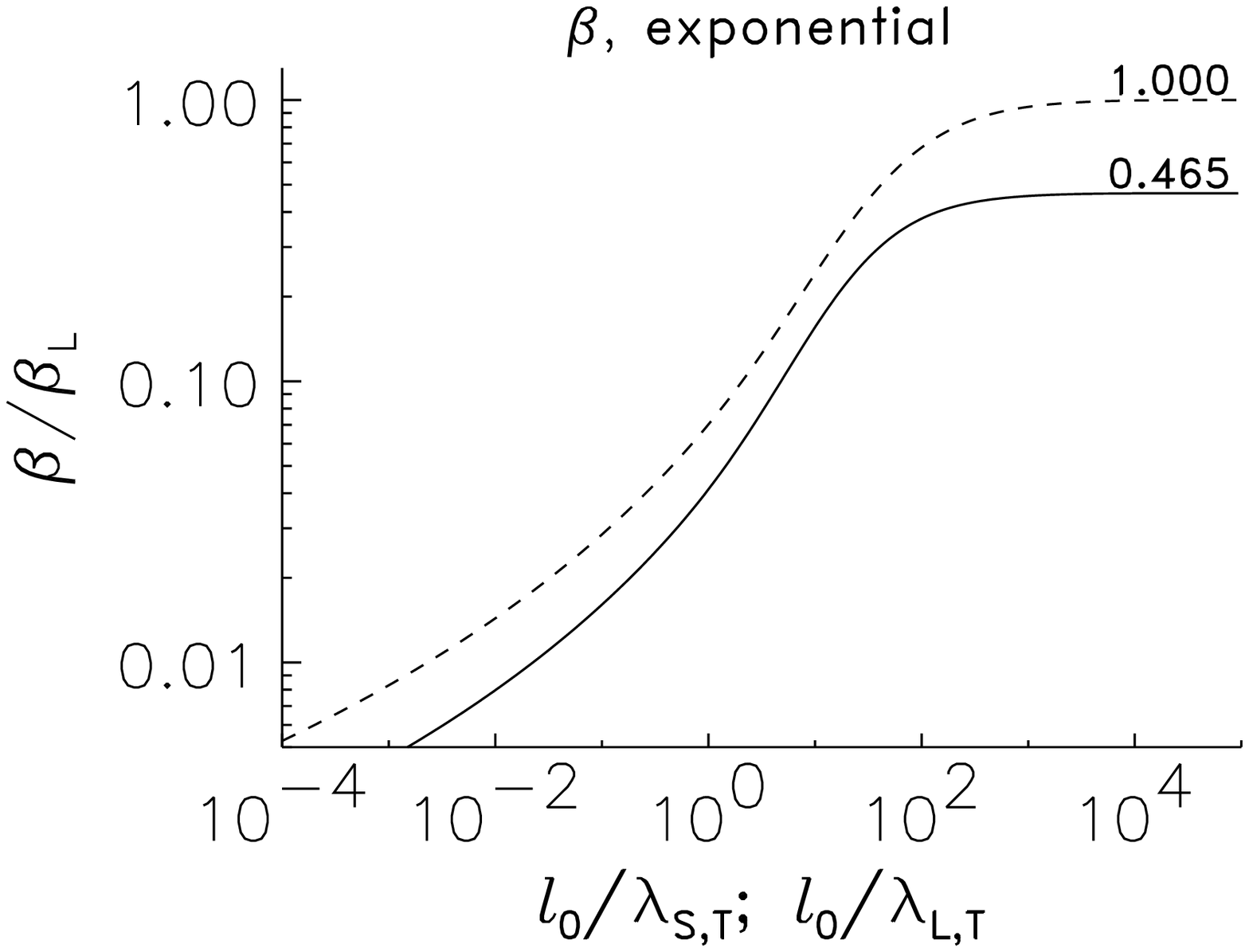}
\includegraphics{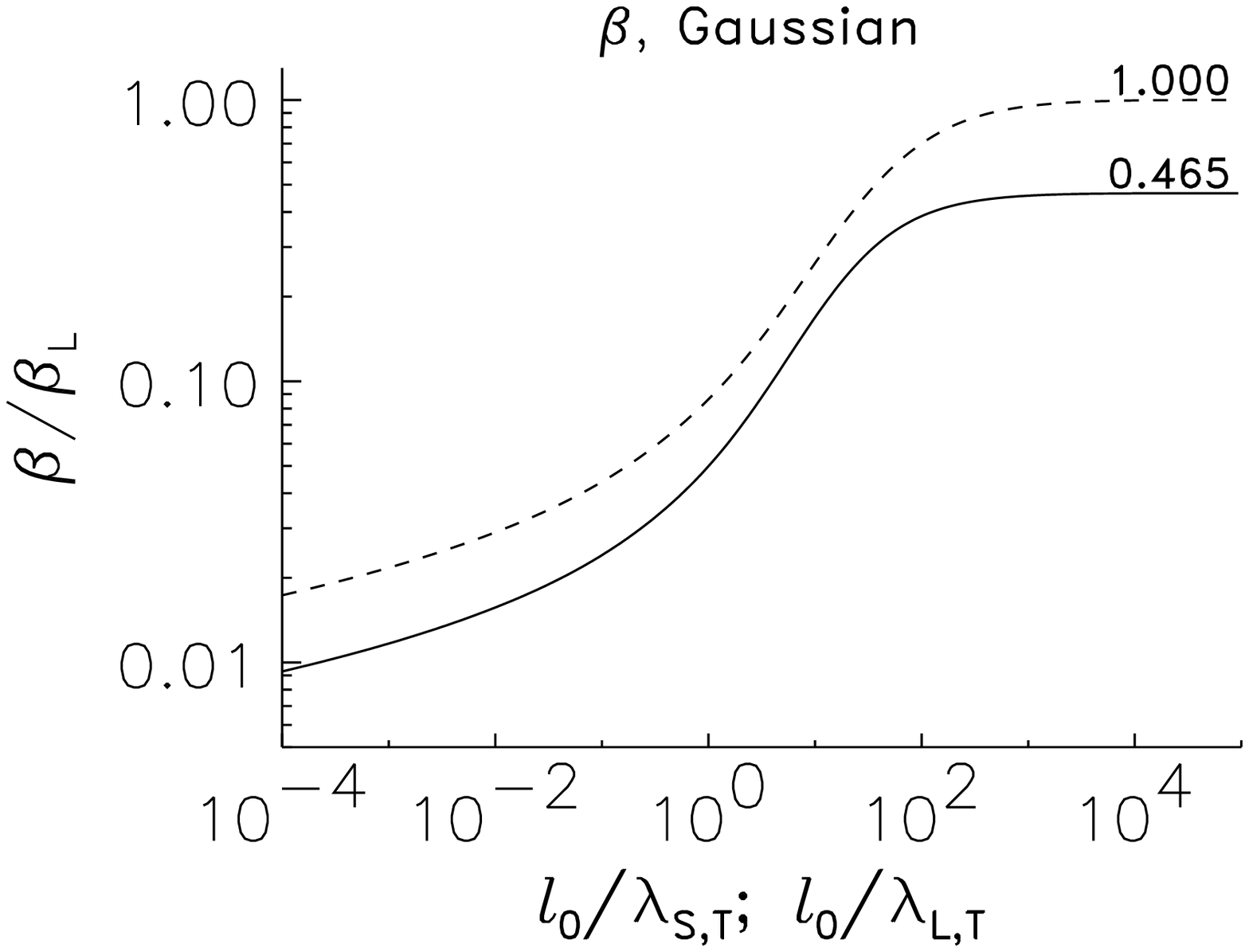}
\includegraphics{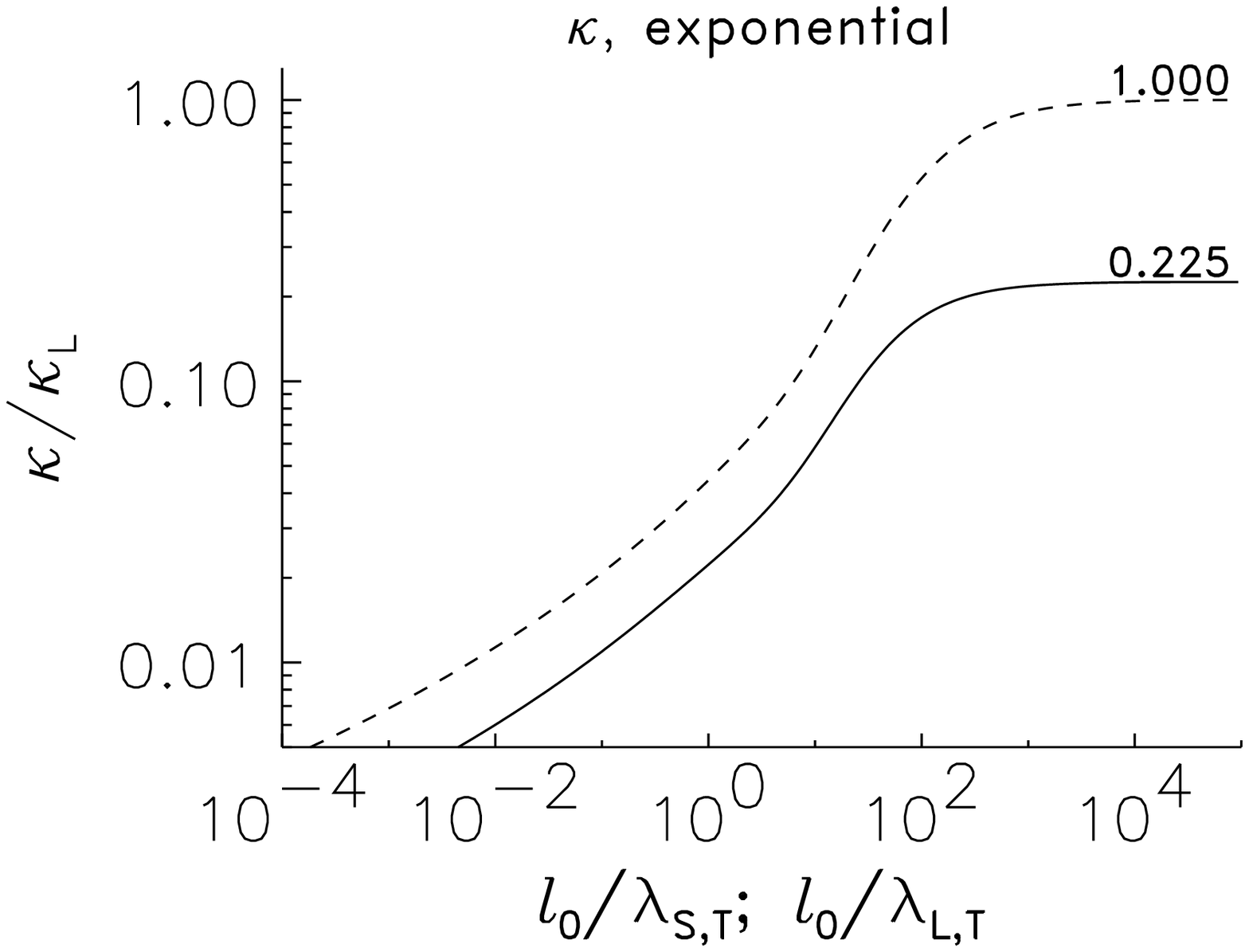}
\includegraphics{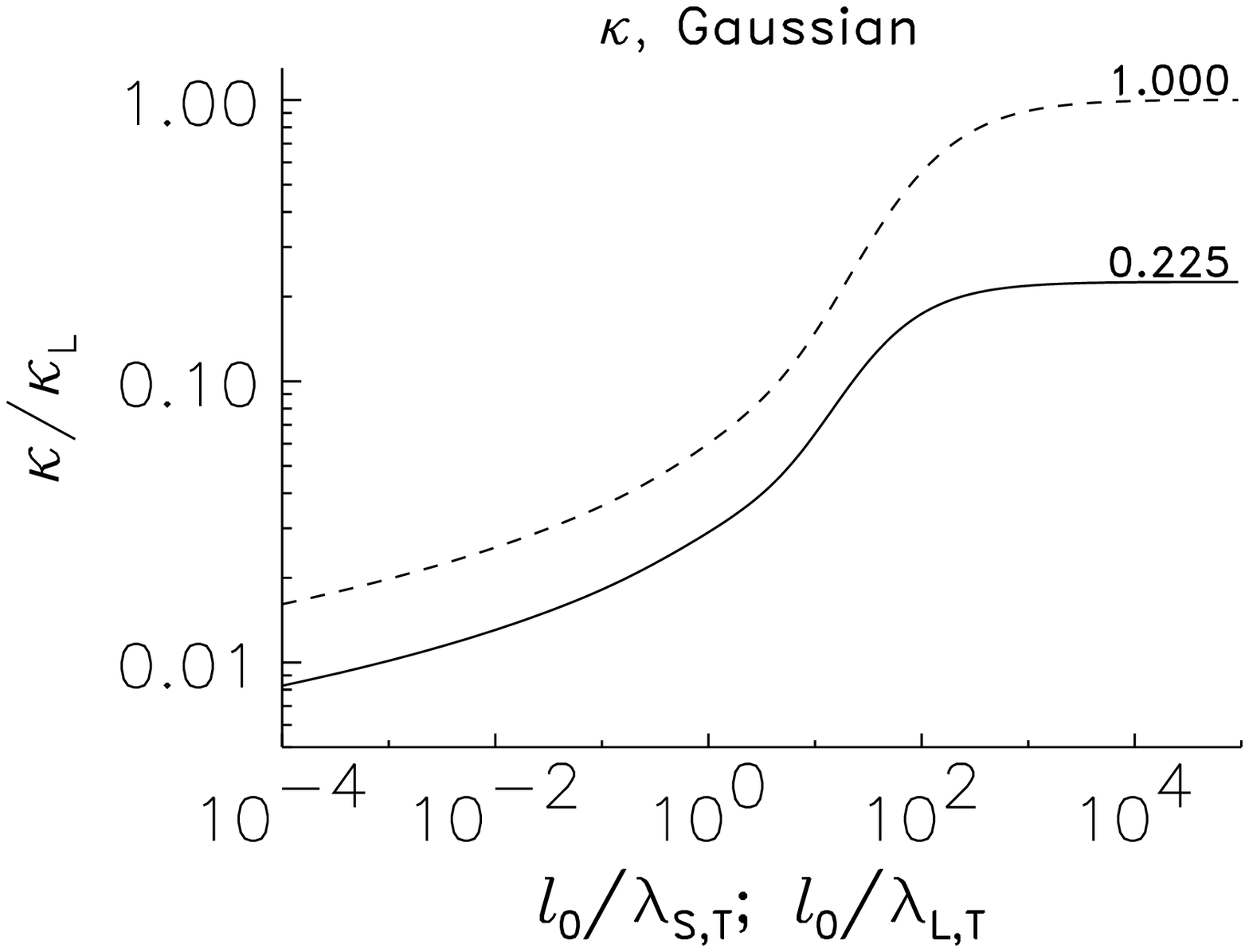}
\caption{Figures on the left/right correspond to the exponential/Gaussian mirror spectra 
[see eqs.~(\ref{SPECTRA})]. 
The solid/dashed lines on all figures show the four transport coefficients, $\sigma$, $\alpha$, $\beta$ 
and $\kappa$, for the Spitzer/Lorentz gas in the presence of stochastic magnetic mirrors as functions of 
the ratio of the magnetic field decorrelation length $l_0$ to the Spitzer/Lorentz electron mean free path, 
$\lambda_{{\rm S},T}$/$\lambda_{{\rm L},T}$, calculated at the electron thermal speed $V_T=\sqrt{2kT/m_e}$
[see eqs.~(\ref{LAMBDA_L_T}),~(\ref{LAMBDA_S_T})].
All transport coefficients are normalized to the standard Lorentz transport coefficients given by 
equations~(\ref{LORENTZ_TRANSPORT}).
The asymptotic values of the coefficients at $l_0/\lambda_T\gg 1$ are given by the numbers 
on the lines. They agree with the results of Spitzer \& H$\ddot{\rm a}$rm (1953).}
\label{FIG_TRANSPORT}
\end{figure}

Using equations~(\ref{LORENTZ_S}) and~(\ref{R_D_LORENTZ}) with $R_{\rm D}$ given in Section~\ref{DIFFUSION},
and numerically performing the velocity integrals, we find all four transport 
coefficients~(\ref{TRANSPORT_COEFFICIENTS}) normalized to the standard Lorentz coefficients~(\ref{LORENTZ_TRANSPORT}). 
The dashed lines in Figures~\ref{FIG_TRANSPORT}(a)--(h) show the resulting normalized transport coefficients
$\sigma$, $\alpha$, $\beta$ and $\kappa$ as functions of $l_0/\lambda_{{\rm L},T}$ for the two mirror spectra: 
(a) exponential, and (b) Gaussian [see equations~(\ref{SPECTRA})]. The asymptotic values of the 
coefficients at large values of $l_0/\lambda_{{\rm L},T}$ are given by the numbers on the dashed lines, and
they are unity. Thus, there is no reductions of the transport coefficients at $l_0/\lambda_{{\rm L},T}\gg 1$, 
as one can expect because there is no reduction of electron diffusivity in this limit [see equation~(\ref{D/D_0})].

In a steady state, the electrical current $j$ in a highly ionized plasma should be zero. Thus, if a 
temperature gradient is present, the resulting electric field $E$ is obtained by setting $j$ to zero
in equation~(\ref{J}). Substituting this result for $E$ into equation for the heat 
flow~(\ref{Q}), we find for the effective thermal conductivity
\beq
\kappa_{\rm eff}&=&\kappa-\alpha\beta/\sigma,
\nonumber\\
\kappa_{\rm eff}/\kappa_{\mbox{\tiny L}}&=&\kappa/\kappa_{\mbox{\tiny L}}-
(3/5)(\alpha/\alpha_{\mbox{\tiny L}})(\beta/\beta_{\mbox{\tiny L}})(\sigma_{\mbox{\tiny L}}/\sigma),
\label{KAPPA_EFF}
\eeq
where we use formulas~(\ref{LORENTZ_TRANSPORT}) for the Lorentz transport coefficients in the 
second line of this equation.

\begin{figure}[p]
\vspace{6.7cm}
\includegraphics{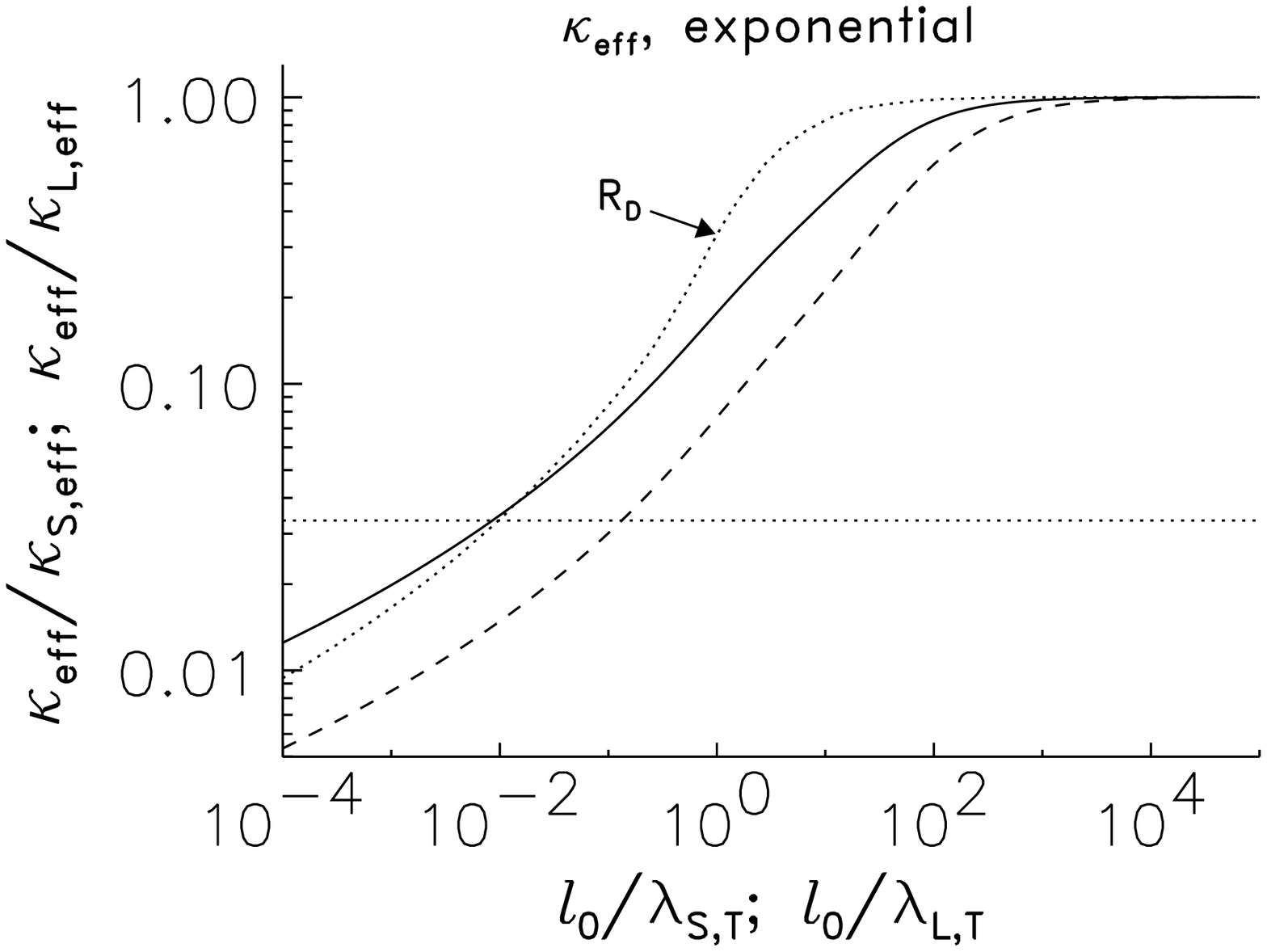}
\includegraphics{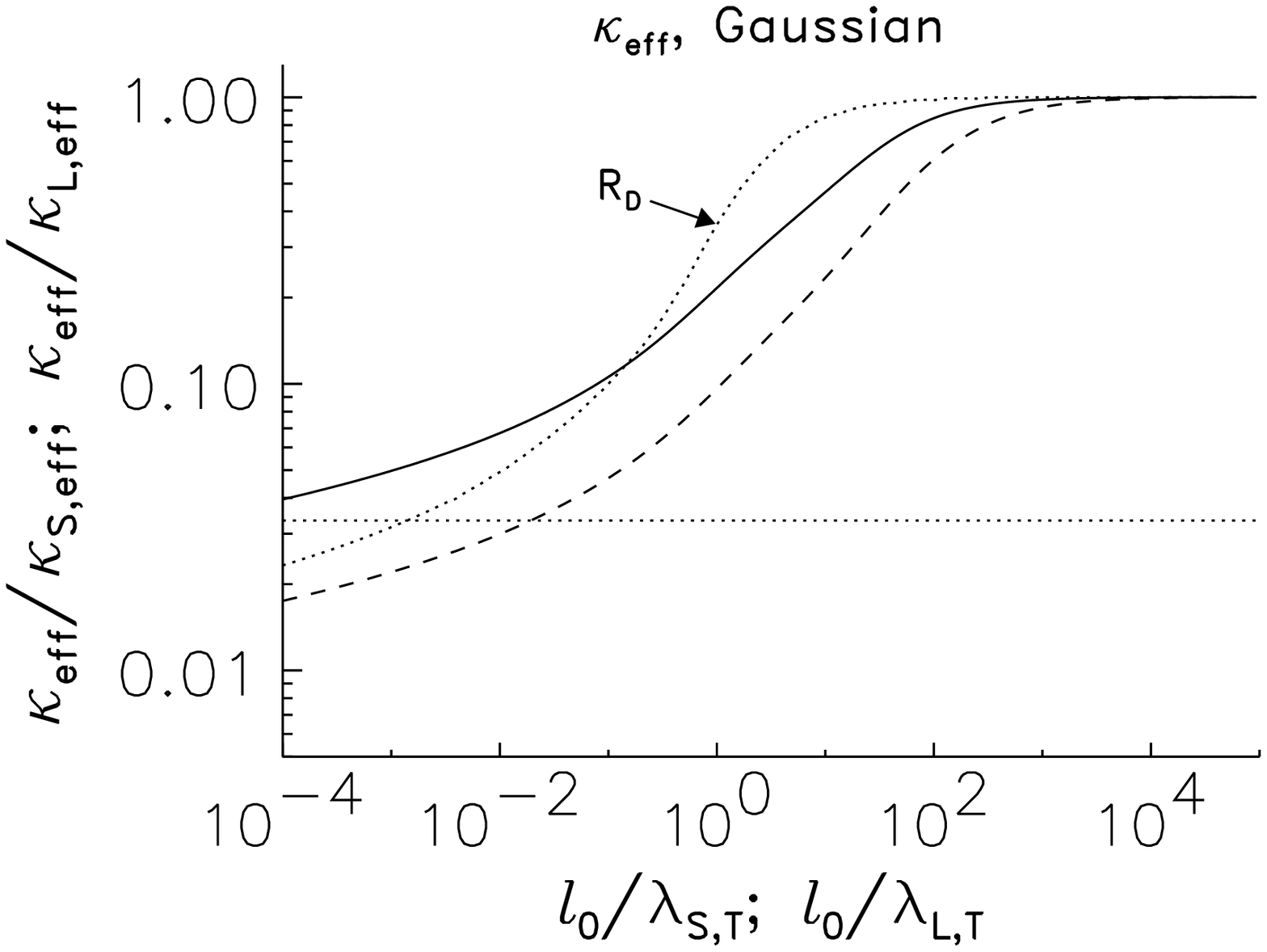}
\caption{The solid/dashed lines show the reduction of the parallel effective thermal conductivity for 
the Spitzer/Lorentz gas by stochastic magnetic mirrors, as function of the ratio of the magnetic field 
decorrelation length to the Spitzer/Lorentz electron mean free path. The notations are the same as in 
Figure~\ref{FIG_TRANSPORT}.
For comparison, we give the mono-energetic diffusion reduction $R_{\rm D}(l_0/\lambda_T)=D(l_0/\lambda_T)/D_0$ 
by the dotted lines.
The horizontal dotted lines represent the reduction of $1/30$, below these lines the thermal conduction is 
so weak, that it should become negligible in clusters of galaxies.}
\label{FIG_CONDUCTION}
\end{figure}

Using the transport coefficients reported in Figures~\ref{FIG_TRANSPORT} by dashed lines and 
formula~(\ref{KAPPA_EFF}), it is easy to find the effective thermal conductivity $\kappa_{\rm eff}$ 
normalized to the standard Lorentz thermal conductivity $\kappa_{\mbox{\tiny L}}$ [see 
equation~(\ref{LORENTZ_TRANSPORT})].
However, it is more useful to give the ratio of $\kappa_{\rm eff}$ to the Lorentz effective 
conductivity, $\kappa_{{\mbox{\tiny L}},{\rm eff}}=0.4\,\kappa_{\mbox{\tiny L}}$. 
This ratio is the actual suppression of the effective conductivity of the Lorentz gas by magnetic mirrors. 
The dashed lines in Figures~\ref{FIG_CONDUCTION} show this suppression, 
$\kappa_{\rm eff}/\kappa_{{\mbox{\tiny L}},{\rm eff}}$, as functions of $l_0/\lambda_{{\rm L},T}$ for 
the two mirror spectra: (a) exponential, and (b) Gaussian [see equations~(\ref{SPECTRA})]. 
It has been estimated that the time of heat conduction in clusters of galaxies is possibly larger 
than the Hubble time if the thermal conductivity is less than $1/30$ of the Spitzer value 
(Suginohara \& Ostriker 1998). The horizontal dotted lines indicate this reduction of $1/30$. 

For comparison, the dotted lines represent the mono-energetic diffusion reduction at the electron 
thermal speed, $R_{\rm D}(l_0/\lambda_{{\rm L},T})=D(l_0/\lambda_{{\rm L},T})/D_0$. We see, that the 
Lorentz gas effective conductivity is reduced to a value two to three times smaller than that of
the diffusion reduction. This is because heat is mainly transported by superthermal electrons. 
These electrons have long mean free paths, and the magnetic mirrors more strongly inhibit their diffusion.

\begin{figure}[p]
\vspace{6.7cm}
\includegraphics{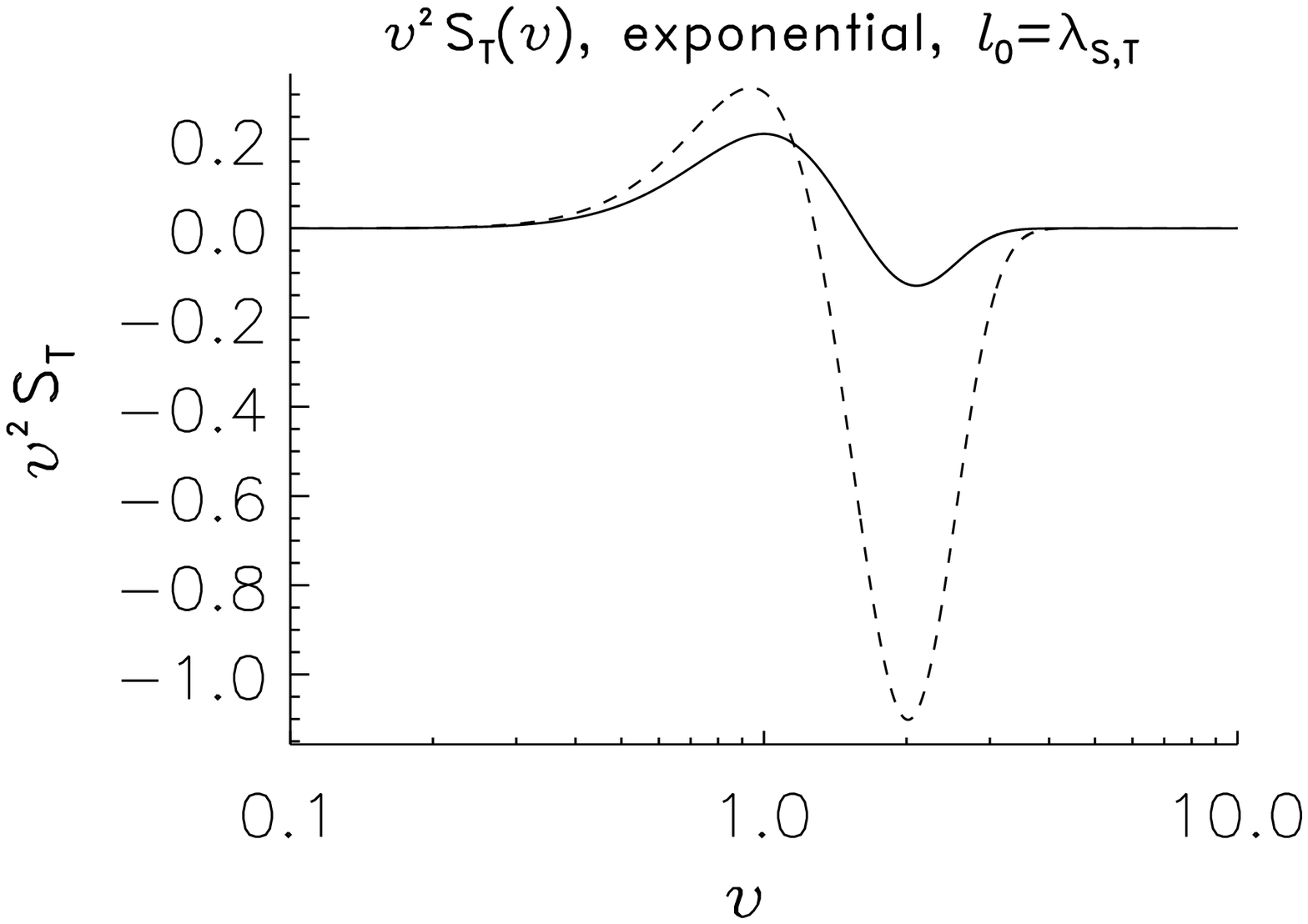}
\includegraphics{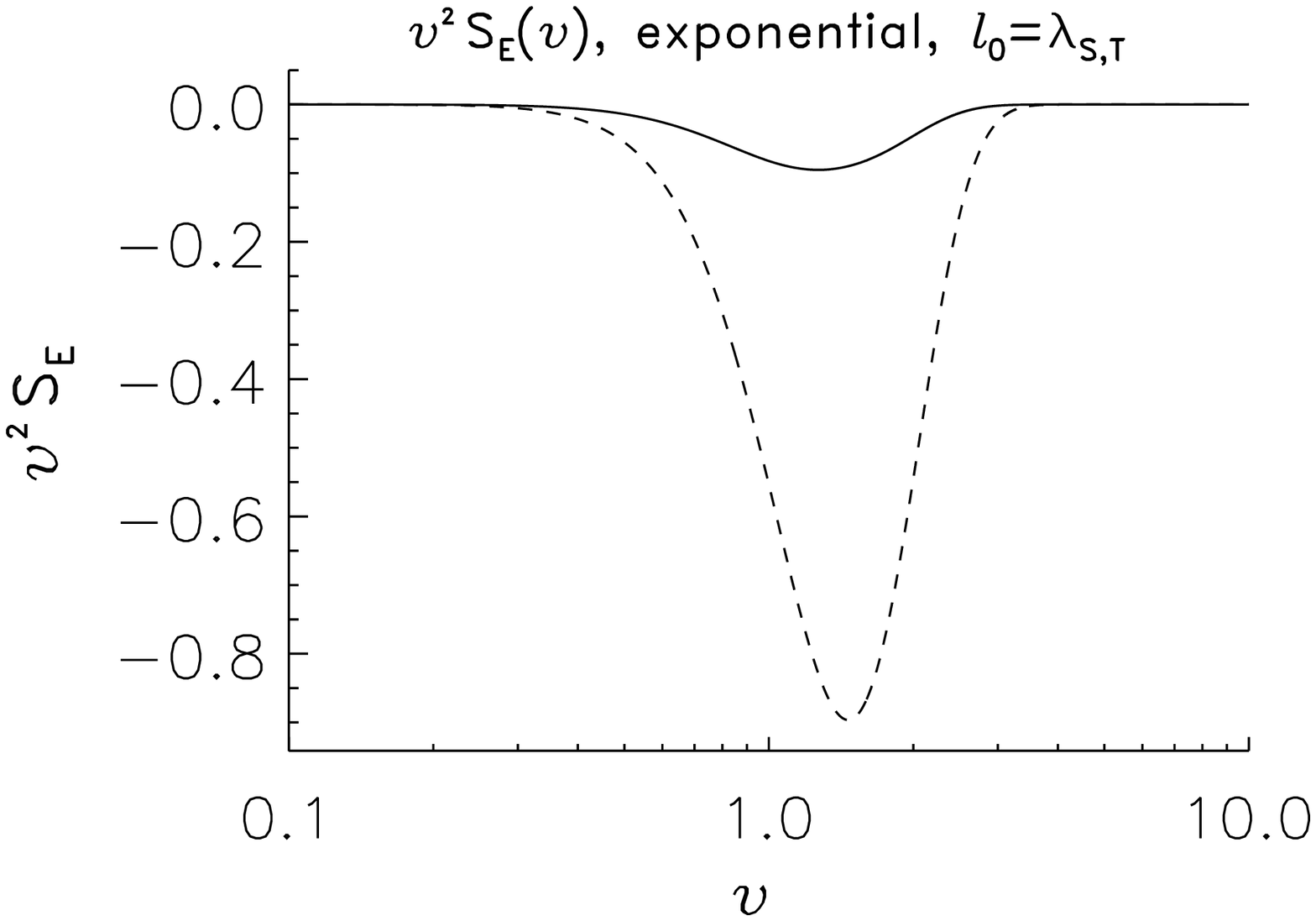}
\caption{The solid lines show functions $\upsilon^2S_T(\upsilon)$ [the left figure] and 
$\upsilon^2S_E(\upsilon)$ [the right figure] for the Spitzer gas in a system of random magnetic mirrors for
the case $l_0=\lambda_{{\rm S},T}$ [$l_0$ is the magnetic field decorrelation length, $\lambda_{{\rm S},T}$ is 
the Spitzer electron mean free path~(\ref{LAMBDA_S_T})]. The dashed lines in the corresponding plots show 
the same functions for the Spitzer gas without mirrors. Both graphs are plotted for the exponential 
mirror spectrum (for the Gaussian spectrum the results are similar).}
\label{FIG_ST_SE}
\end{figure}


\subsection{Spitzer gas in a system of random mirrors}\label{SPITZER_GAS}

Now consider the full collision integral~(\ref{COLLISION_INTEGRAL}) for the Spitzer gas in a system 
of random magnetic mirrors. We have numerically solved the full set of our 
equations~(\ref{L_OPERATOR})--(\ref{BOUNDARY_CONDITIONS}). 
The formulas~(\ref{TRANSPORT_COEFFICIENTS}) and~(\ref{KAPPA_EFF}) remain the same as for the Lorentz gas, 
but the functions $S_T(\upsilon)$ and $S_E(\upsilon)$ are different. For Spitzer electrons the mean 
free path is $\lambda_{\rm S}\propto V^4{[1+\Phi(\upsilon)-G(\upsilon)]}^{-1}$ [Spitzer 1962, the error 
functions $\Phi$ and $G$ are given by~(\ref{PHI_G})]. Thus, formula~(\ref{R_D_LORENTZ}) for the reduction 
of spatial diffusivity now becomes
\beq
R_{\rm D}=R_{\rm D}\left(l_0/\lambda_{\rm S}\vphantom{S'}\right)=R_{\rm D}\left(
\upsilon^{-4}\frac{l_0}{\lambda_{{\rm S},T}}\frac{1+\Phi(\upsilon)-G(\upsilon)}{1+\Phi(1)-G(1)}\right),
\label{R_D_SPITZER}
\eeq
where the Spitzer electron mean free at the thermal speed $V_T=\sqrt{2kT/m_e}$ is
\beq
\lambda_{{\rm S},T}=0.614\left.(kT)^2\right/\pi n e^4\ln\Lambda
\approx 0.06\,{\rm Kpc}\,{(T/10^7{\rm K})}^2(10^{-3}{\rm cm}^{-3}/n).
\label{LAMBDA_S_T}
\eeq

Functions $S_T(\upsilon)$ and $S_E(\upsilon)$ are defined by equations~(\ref{S_E_T}), and they are
the two inhomogeneous solutions of the equations~(\ref{KINETIC_EQUATION}),~(\ref{BOUNDARY_CONDITIONS}).
To find these solutions we solved equation~(\ref{KINETIC_EQUATION}) numerically by iterations. At each 
iteration step the integral part of this equation, ${\hat{\cal I}}S$, was calculated using the solution 
for $S$ from the previous step, and the new solution for $S$ was calculated by the Gaussian decomposition 
with backsubstitution (Fedorenko 1994), using the boundary conditions~(\ref{BOUNDARY_CONDITIONS}). 
Initially, we started with zero function $S=0$. The iterations converged very rapidly, and the Gaussian 
decomposition method is stable. We believe that our numerical method is much better and faster than the 
method of Spitzer and H$\ddot{\rm a}$rm (1953) because their method was not stable. It took us less 
than ten seconds of computer time to calculate all digits of the transport coefficients reported by 
Spitzer and H$\ddot{\rm a}$rm.

The solid lines in Figures~\ref{FIG_TRANSPORT}(a)--(h) show the resulting transport coefficients
$\sigma$, $\alpha$, $\beta$ and $\kappa$ normalized to the standard Lorentz 
coefficients~(\ref{LORENTZ_TRANSPORT}) as functions of $l_0/\lambda_{{\rm S},T}$ for the two mirror spectra: 
(a) exponential, and (b) Gaussian [see equations~(\ref{SPECTRA}); remember that $l_0$ is the magnetic field 
decorrelation length]. The asymptotic values of the coefficients at large values of $l_0/\lambda_{{\rm S},T}$ 
are given by the numbers on the solid lines, and they agree with the results of Spitzer and H$\ddot{\rm a}$rm. 

The effective thermal conductivity, $\kappa_{\rm eff}$, normalized to the Spitzer effective 
conductivity, $\kappa_{{\mbox{\tiny S}},{\rm eff}}=0.0943\,\kappa_{\mbox{\tiny L}}$, is given in 
Figures~\ref{FIG_CONDUCTION} by the solid lines for the two mirror spectra. This normalized conductivity is 
the actual suppression of the effective thermal conductivity of the Spitzer gas by stochastic magnetic mirrors. 
{\it It is the result that should be applied in astrophysical problems with random magnetic mirrors.}

Finally, it is interesting to see how the mirrors change the Spitzer perturbed electron distribution
function. In Figure~\ref{FIG_ST_SE} we plot functions $\upsilon^2S_T(\upsilon)$ and $\upsilon^2S_E(\upsilon)$ 
for the case when $l_0=\lambda_{{\rm S},T}$ [note that $2\pi V^2f_1$ is the actual distribution of electrons 
over speed $V=\upsilon V_T$, see equation~(\ref{F_1_GENERAL})]. 
The solid lines represent these functions for the Spitzer gas in a system of random mirrors 
with the exponential mirror spectrum (for the Gaussian spectrum the results are similar). The dashed lines
show the same functions for the Spitzer gas without magnetic mirrors.
We see that $\upsilon^2S_T(\upsilon)$ and $\upsilon^2S_E(\upsilon)$ are reduced at large values of 
$\upsilon$, i.e.~magnetic mirrors reduce the anisotropy of the superthermal electrons, which carry the 
electrical current and heat.


\section{Conclusions}\label{CONCLUSIONS}

In this paper we have derived the actual parallel effective thermal conductivity that should be applied
to astrophysical systems with random magnetic mirrors, as well as other important transport coefficients.

Now, let us apply our results for the reduction of the Spitzer effective electron thermal conductivity, 
shown in Figure~\ref{FIG_CONDUCTION} by the solid lines, to the galaxy cluster formation problem.
If the reduction is by more than a factor of thirty (shown by the horizontal dotted lines in 
Figure~\ref{FIG_CONDUCTION}), then the time of heat transport becomes larger than the Hubble time, and the 
heat conduction can be neglected (Suginohara \& Ostriker 1998).~\footnote{See the footnote on page~\pageref{FOOTNOTE}.}
We see that this is the case if the magnetic field decorrelation length $l_0$ is roughly less than 
$10^{-4}\,$--$\,10^{-2}$ of the electron mean free path at the thermal speed $\lambda_T=\lambda(\sqrt{2kT/m_e})$ 
(we consider the Spitzer gas).
Although there is little observational data about the topology of magnetic fields in clusters of galaxy, 
the magnetic field scale is probably $1\,$--$\,10\,{\rm Kpc}$ (Kronberg 1994; Eilek 1999 and 
references in it). According to equation~(\ref{LAMBDA_S_T}) the characteristic electron mean free path 
at the thermal speed is $0.06\,$--$\,60\,{\rm Kpc}$ for temperatures $T=10^7\,$--$\,10^8\,{\rm K}$ and
densities $n=10^{-4}\,$--$\,10^{-3}\,{\rm cm}^{-3}$.
We see, that in general, the effective electron thermal conductivity parallel to the magnetic field lines is
not reduced enough by magnetic mirrors to be completely neglected. However, as we pointed out in the 
introductory section, there is an additional effect that electrons have to travel along tangled 
magnetic field lines larger distances from hot to cold regions of space, so the thermal conduction 
is further reduced (this effect will be considered in our future paper). 
At the moment, ``whether electron thermal conductivity in clusters of galaxies is sufficiently inhibited 
that it can be ignored'' is still an open question.

Recently, Cowley, Chandran {\it et al.} studied the reduction of the parallel thermal conduction, and 
they concluded that the thermal conductivity in galaxy clusters is reduced enough to be neglected 
(Chandran \& Cowley 1998; Chandran, Cowley \& Ivanushkina 1999; Albright {\it et al.}~2000). Their 
conclusions are different from ours. The reason is that our approach in calculation of the conductivity 
is very different, and our results are qualitatively different. 
The main difference is that they took the reduction of thermal conductivity to be equal to the reduction 
of diffusivity of thermal electrons. In fact, the reduction of diffusivity is due to the enhanced pitch 
angle scattering by stochastic magnetic mirrors, and to find the reduction of thermal conductivity, the 
full set of kinetic equations must be derived and solved. This consistent way of solving the 
problem makes a considerable difference (see Figure~\ref{FIG_CONDUCTION}). On the other hand, 
Cowley, Chandran and {\it et al.} first called attention to the importance of the effective mean free path 
$\lambda_{\rm eff}$ and found the correct qualitative result, that in the limit $l_0\ll\lambda$ the 
diffusion reduction is controlled by the mirrors whose spacing is of order of the effective mean free.

\acknowledgments

We are happy to acknowledge many useful discussions of this problem with Jeremiah Ostriker, Jeremy Goodman 
and David Spergel. We would also like to thank Makoto Matsumoto, Takuji Nishimura and Shawn J. Cokus 
providing us with fast random number generators (which are given at http://www.math.keio.ac.jp/matumoto/emt.html).
This work was partially supported by DOE under Contract No. DE-AC 02-76-CHO-3073. Leonid Malyshkin
would also like to thank the Department of Astrophysical Sciences at Princeton University for financial support.


\appendix

\section{Solution of equation~(\ref{KINETIC_EQ}) in the limit {\boldmath $l_m\ll\lambda_{\rm eff}$}}\label{A_CASE_1}

Here we solve equation~(\ref{KINETIC_EQ}) by expansion in the limit $l_m\ll \lambda_{\rm eff}$. This condition 
means that collisions are too weak to scatter the electron out of the loss cones. Therefore, $F(x,\mu)\equiv 0$ 
when $|\mu|>\mu_{\rm crit}$. 
 
We make use the fact that $(V/\nu)(\partial/\partial x)\sim \lambda/l_m\gg 1$. Also we will show that 
$1/\nu\tau_m\ll 1$. The validity of this last assumption appears below. To zero order, we have 
$\partial F/\partial x=0$, and $F(x,\mu)=F_0(\mu)$. $F_0(-\mu)=F_0(\mu)$ because of electron reflection 
at the mirrors and the symmetry of the loss cones.
Up to first order, $F(x,\mu)=F_0(\mu)+F_1(x,\mu)$, and we have
\beq
\mu V\frac{\partial F_1}{\partial x}=\frac{\nu}{2}
\frac{\partial}{\partial\mu}\left[(1-\mu^2)\frac{\partial F_0}{\partial\mu}\right]+\frac{F_0}{\tau_m}.
\eeq
We integrate this equation over $x$ along a closed back and forth trajectory of a trapped electron 
shown by the dotted lines in Figure~\ref{FIG_MIRRORS}(b), to obtain
\beq
\left\{
\begin{array}{l}
\partial/\partial\mu\left[(1-\mu^2)\partial F_0/\partial\mu\right]+2F_0/\nu\tau_m=0,\\
F_0(-\mu)=F_0(\mu), \quad F_0(\pm\mu_{\rm crit})=0.
\end{array}
\right.
\label{A_CASE_1_F0}
\eeq

We solve equation~(\ref{A_CASE_1_F0}) by a further expansion, $1/\nu\tau_m\ll 1$.
The even solution in the ``inside'' region $1-|\mu|\gg e^{-\nu\tau_m}$ up to first order is
\beq
F_0^{(i)}=C^{(i)}\left[1-\frac{1}{\nu\tau_m}\ln{\frac{1}{1-\mu^2}}\right], \quad 1-|\mu|\gg e^{-\nu\tau_m}.
\label{A_CASE_1_I}
\eeq
On the other hand, the zero order solution in the ``boundary'' regions $1-|\mu|\ll 1$ is
\beq
F_0^{(b)}&=&C^{(b)}\ln{\frac{1-|\mu|}{1-\mu_{\rm crit}}}, \quad 1-\mu_{\rm crit}\le 1-|\mu|\ll 1. 
\label{A_CASE_1_B}
\eeq
We match solutions~(\ref{A_CASE_1_I}) and~(\ref{A_CASE_1_B}) together in regions 
$e^{-\nu\tau_m}\ll 1-|\mu|\ll 1$ to finally obtain $\tau_m=\nu^{-1}\ln m$, justifying $1/\nu\tau_m\ll 1$.
This is the first result in equation~(\ref{TAU_CASE}).


\section{Solution of equation~(\ref{KINETIC_EQ}) in the limits 
{\boldmath $\lambda_{\rm eff}\ll l_m\ll\lambda^2/\lambda_{\rm eff}$} and 
{\boldmath $\lambda^2/\lambda_{\rm eff}\ll l_m$}}\label{A_CASES_2_3}

Let us consider the kinetic equation~(\ref{KINETIC_EQ}) in the more limited case $\lambda\ll l_m$ 
(note that $\lambda_{\rm eff}\ll\lambda$). This means that in the kinetic equation 
$(V/\nu)(\partial/\partial x)\simlt\lambda/l_m\ll 1$. We will also show that $1/\nu\tau_m\ll 1$. The validity of 
this assumption appears below. To zero order, we have $\partial F/\partial\mu=0$, so $F(x,\mu)=F_0(x)$. 
$F_0(-x)=F_0(x)$ because of symmetry. Up to first order, $F(x,\mu)=F_0(x)+F_1(x,\mu)$, and we have
\beq
\frac{\nu}{2}\frac{\partial}{\partial\mu}\left[(1-\mu^2)\frac{\partial F_1}{\partial\mu}\right]=
\mu V\frac{\partial F_0}{\partial x}-\frac{F_0}{\tau_m}.
\eeq
We integrate the above equation over $\mu$, and then set $\mu=\pm 1$ to find the constant of integration. 
As a result, we obtain $F_0/\tau_m\ll V(\partial F_0/\partial x)$ [so, $1/\nu\tau_m$ is of second order], and 
$\partial F_1/\partial\mu=-(V/\nu)(\partial F_0/\partial x)$. We integrate this last equation over $\mu$ once 
more, and obtain
\beq
F_1=-(\mu V/\nu)(\partial F_0/\partial x)+C(x),
\label{A_CASES_2_3_F_1}
\eeq
where $C(x)$ is another integration constant. 

We continue the expansion of the kinetic equation~(\ref{KINETIC_EQ}) to next order. Up to second order, 
$F(x,\mu)=F_0(x)+F_1(x,\mu)+F_2(x,\mu)$. Using equation~(\ref{A_CASES_2_3_F_1}), we have
\beq
\frac{\nu}{2}\frac{\partial}{\partial\mu}\left[(1-\mu^2)\frac{\partial F_2}{\partial\mu}\right]=
-\frac{\mu^2V^2}{\nu}\frac{\partial^2 F_0}{\partial^2 x}+\mu V\frac{\partial C}{\partial x}-\frac{F_0}{\tau_m}.
\label{A_CASES_2_3_F_2}
\eeq
We integrate equation~(\ref{A_CASES_2_3_F_2}) over $\mu$ from $-1$ to $1$ and obtain
\beq
\left\{
\begin{array}{l}
\partial^2 F_0/\partial^2 x+(3\nu/\tau_m V^2) F_0=0,\\
F_0(-x)=F_0(x).
\end{array}
\right.
\eeq
Finally, we integrate this equation and obtain zero order solution for the time-dependent distribution 
function~(\ref{F_TIME_DEPENDENT})
\beq
f(t,x)=e^{-t/\tau_m} F_0(x)=e^{-t/\tau_m}\cos{\left(x\sqrt{3\nu/\tau_m V^2}\,\right)},
\label{A_CASES_2_3_F}
\eeq
where we drop an unnecessary normalization constant of integration.

Now, to find $\tau_m$, we calculate the flux of escaping electrons through the two escape windows 
(see Figure~\ref{FIG_MIRRORS})
\beq
\partial N/\partial t=-2\int_{\mu_{\rm crit}}^1 \mu V f(t,l_m/2)\,d\mu=
-(V/m)\,e^{-t/\tau_m}\cos{\left[(l_m/2)\sqrt{3\nu/\tau_m V^2}\,\right]},
\label{A_CASES_2_3_FLUX_1}
\eeq
where we use equation~(\ref{A_CASES_2_3_F}) for $f(t,l_m/2)$.
On the other hand, the flux is equal to the change of the total number of electrons
\beq
\partial N/\partial t=\int_{-l_m/2}^{l_m/2} \int_{-1}^1 (\partial f/\partial t)\,d\mu\,dx=
-(4/\tau_m)\,e^{-t/\tau_m}\sqrt{\tau_m V^2/3\nu}\:\sin{\left[(l_m/2)\sqrt{3\nu/\tau_m V^2}\,\right]}.
\label{A_CASES_2_3_FLUX_2}
\eeq
Equating the two formulas for $\partial N/\partial t$, we obtain
\beq
(3/16)(\nu\tau_m/m^2)=\tan^2{\left(\sqrt{3\nu l_m^2/4\tau_m V^2}\right)}.
\label{A_CASES_2_3_TAU}
\eeq

In the limit $\lambda\ll l_m\ll\lambda^2/\lambda_{\rm eff}$ the argument of the tangent above is small,
so we expand the tangent and obtain $\tau_m=\nu^{-1} (l_m/\lambda_{\rm eff})$, while $F_0\approx {\rm const}$. 
In the limit $\lambda^2/\lambda_{\rm eff}\ll l_m$ the left hand side of equation~(\ref{A_CASES_2_3_TAU}) is 
large, therefore, the argument of the tangent is $\pi/2$, and we have 
$\tau_m=\nu^{-1} (3/\pi^2){(l_m/\lambda)}^2$ [the third line in equation~(\ref{TAU_CASE})], i.e. the escape 
time is controlled by diffusion in $x$-space. In both limits $1/\nu\tau_m\ll 1$, as we assumed above, 
(and of second order).

Now, the limit $\lambda_{\rm eff}\ll l_m\simlt\lambda$ is still left. The result in this case is the 
same as the result in case $\lambda\ll l_m\ll\lambda^2/\lambda_{\rm eff}$. However, instead of solving the 
kinetic equation, we give the following qualitative arguments supported by our numerical simulations 
(see Figure~\ref{FIG_TAU}). The relaxation time of the electron distribution in $\mu$-space can be estimated 
as ${\Delta t}_\mu\sim \nu^{-1}$. The relaxation time in $x$-space can be estimated as the crossing time
${\Delta t}_x\sim l_m/V=\nu^{-1}(l_m/\lambda)$ in case $l_m\simlt\lambda$, and as the time of diffusion
across ${\Delta t}_x\sim \nu^{-1}{(l_m/\lambda)}^2$ in case $\lambda\ll l_m$. All relaxation times are small 
compared to the escape time $\tau_m$, i.e.~${\Delta t}_\mu,{\Delta t}_x\ll\tau_m$ for the entire range
$\lambda_{\rm eff}\ll l_m\ll\lambda^2/\lambda_{\rm eff}$. This means that the distribution function is 
approximately constant in $x$ and $\mu$, say $F_0\approx 1$, $f\approx e^{-t/\tau_m}$. We then carry out 
calculations similar to those we used in formulas~(\ref{A_CASES_2_3_FLUX_1}) and~(\ref{A_CASES_2_3_FLUX_2}) 
to find that $\tau_m=\nu^{-1} (l_m/\lambda_{\rm eff})$ [the second line in equation~(\ref{TAU_CASE})].


\section{The Additional electron flow produced by electric field}\label{E_FLUXES}

Let us, for simplicity, consider the Lorentz gas. The results for the Spitzer gas are similar.

First, we derive an estimate for the additional flow $d{\tilde{\cal F}}$ of electrons that are in an 
interval $V\in[V,V+dV)$ of the velocity space, produced by an electric field $E$ due to the change 
of the two loss cones of a mirror trap. Let us consider only the principle mirrors, because they 
mainly control the diffusion of electrons (see Section~\ref{DIFFUSION}). In this appendix, we denote 
their mirror strength (the principle mirror strength) as $M$.

The principle mirror strength is of order of five, so in the case when the magnetic field decorrelation 
length is more than or approximately equal to the electron mean free path, $l_0\simgt\lambda$, the 
escape of electrons from the mirror trap is mainly controlled by their spatial diffusion, see 
Section~\ref{ESCAPE}. Thus, in this limit, the electrons ``do not care'' about the size of the 
loss cones, and therefore, no additional flow arises.

In the case $l_0\simlt\lambda$ there is a non-zero additional flow $d{\tilde{\cal F}}$. 
In Figure~\ref{FIG_MIRRORS}(b), because of the electric field, the loss cone on the left, 
$\mu_{{\rm crit},-}$, is not equal to that on the right, $\mu_{{\rm crit},+}$. The size of the two loss 
cones is estimated from the conservation of the electron magnetic moment, $(1-\mu^2)V^2/B={\rm const}$, 
and from the conservation of energy, $m_eV^2/2+eEx={\rm const}$. We have
\beq
\mu_{{\rm crit},\pm}^2\approx 1-1/M\pm(eEl_M/m_eV^2M),
\label{A_E_FLUXES_MU}
\eeq
where $l_M$ is the spacing of the principle mirrors.

Let $d{\tilde{\cal F}}_+$ and $d{\tilde{\cal F}}_-$ be the absolute values of the fluxes of the escaping 
electrons to the right and to the left respectively. Then, their sum is
\beq
d{\tilde{\cal F}}_++d{\tilde{\cal F}}_-=(l_M/\tau_M)\:2\pi V^2f_0dV,
\label{A_E_FLUXES_F_SUM}
\eeq
where $2\pi V^2f_0dV$ is the number density of electrons expressed in terms of the Maxwellian zero 
order electron distribution function $f_0$, and $\tau_M$ is the escape time, see 
equations~(\ref{F_TIME_DEPENDENT}),~(\ref{TAU}),~(\ref{MAXWELL}) and Section~\ref{ESCAPE}. The actual 
electron flow, $d{\tilde{\cal F}}$, is equal to the difference of $d{\tilde{\cal F}}_+$ and 
$d{\tilde{\cal F}}_-$, because they are in opposite directions. An estimate for the ratio 
$d{\tilde{\cal F}}/(d{\tilde{\cal F}}_++d{\tilde{\cal F}}_-)$ is
\beq
\frac{d{\tilde{\cal F}}}{d{\tilde{\cal F}}_++d{\tilde{\cal F}}_-}
=\frac{d{\tilde{\cal F}}_+-d{\tilde{\cal F}}_-}{d{\tilde{\cal F}}_++d{\tilde{\cal F}}_-}
\approx\left.\left[\int_{\mu_{{\rm crit},+}}^1\!\!\!\mu d\mu-\int_{\mu_{{\rm crit},-}}^1\!\!\!\mu d\mu\right]
\right/\left[\int_{\mu_{{\rm crit},+}}^1\!\!\!\mu d\mu+\int_{\mu_{{\rm crit},-}}^1\!\!\!\mu d\mu\right].
\label{A_E_FLUXES_RATIO}
\eeq
Now, using equations~(\ref{A_E_FLUXES_MU})--(\ref{A_E_FLUXES_RATIO}), we obtain the additional electron flow
\beq
d{\tilde{\cal F}}\approx -2\pi(eE/m_e)f_0(l_M^2/\tau_M)dV.
\eeq
The factor $(l_M^2/\tau_M)$ in this equation is proportional to the spatial diffusivity of the electrons
(provided that the diffusivity is controlled by the principle mirrors, see Section~\ref{DIFFUSION}).
Thus, it is obviously $l_M^2/\tau_M=R_{\rm D}\,\lambda_{\mbox{\tiny L}}^2\nu_{\mbox{\tiny L}}$,
where $\lambda_{\mbox{\tiny L}}$ and $\nu_{\mbox{\tiny L}}$ are the standard Lorentz mean free path and 
collision frequency, and $R_{\rm D}$ is the reduction of the spatial diffusivity reported
in Section~\ref{DIFFUSION}. Using that $\lambda_{\mbox{\tiny L}}\propto V^4$,
$\nu_{\mbox{\tiny L}}\propto V^{-3}$, and equations~(\ref{MAXWELL}),~(\ref{LAMBDA_L_T}), 
$V_T=\sqrt{2kT/m_e}$, we finally obtain
\beq
d{\tilde{\cal F}}\approx -(1/2){(2/\pi)}^{3/2}\left(\left.k^{3/2}T^{3/2}E\right/m_e^{1/2}e^3\ln\Lambda\right)
R_{\rm D}\,\upsilon^5\,e^{-\upsilon^2} d\upsilon.
\label{A_E_FLUXES_FLUX}
\eeq

Now we like to compare this result for the additional flow $d{\tilde{\cal F}}$ with the main flow 
$d{\cal F}$ produced by the electric field due to acceleration of particles. The later is
\beq
d{\cal F}=\int_{-1}^1\!\mu V\,f_1\,d\mu\,2\pi V^2dV\,d\mu
=-(2/3){(2/\pi)}^{3/2}\left(\left.k^{3/2}T^{3/2}E\right/m_e^{1/2}e^3\ln\Lambda\right)
R_{\rm D}\,\upsilon^7\,e^{-\upsilon^2} d\upsilon,
\eeq
where we substituted function $f_1$ given by~(\ref{F_1}), and function 
$S(\upsilon)=\gamma_{\mbox{\tiny E}}S_E(\upsilon)$ given by~(\ref{GAMMA_T_E}) and~(\ref{LORENTZ_S}). 
As a result,
\beq
d{\tilde{\cal F}}/d{\cal F}\approx (3/4)\,\upsilon^{-2}.
\eeq
Because the electrical current and the heat flow are mainly transported by superthermal electrons 
$\upsilon^2\sim 4$, the additional flow produced by electric field due to non-equal loss cones can indeed 
be neglected in comparison with the main flow due to acceleration of electrons by electric field.



\end{document}